\documentclass{article}
\usepackage[twocolumn]{emulateapj}
\usepackage{epsf}

\def\lae{\mathrel{<\kern-1.0em\lower0.9ex\hbox{$\sim$}}}
\def\gae{\mathrel{>\kern-1.0em\lower0.9ex\hbox{$\sim$}}}

\slugcomment{Accepted for publication in the Astrophysical Journal}

\lefthead{{\sc C\^OT\'E, WEST \& MARZKE}}
\righthead{Globular Clusters and the Missing Satellite Problem}

\begin{document}

\title{Globular Cluster Systems and the Missing Satellite Problem: Implications for Cold Dark Matter Models}

\author{Patrick C\^ot\'e}
\affil{Department of Physics and Astronomy, Rutgers University, New Brunswick, NJ 08854 \\ pcote@physics.rutgers.edu}
\medskip

\author{Michael J. West}
\affil{Department of Physics \& Astronomy, University of Hawaii, Hilo, HI 96720 \\ west@bohr.uhh.hawaii.edu}
\medskip

\and

\author{Ronald O. Marzke}
\smallskip
\affil{Department of Physics \& Astronomy, San Francisco State University, 
1600 Holloway Avenue, San Francisco, CA 94132 \\ marzke@quark.sfsu.edu}

\medskip

\begin{abstract}
We analyze the metallicity distributions of globular clusters belonging to
28 early-type galaxies in the survey of Kundu \& Whitmore (2001). A Monte Carlo
algorithm which simulates the chemical evolution of galaxies that grow hierarchically
via dissipationless mergers is used to determine the most probable protogalactic
mass function for each galaxy.
Contrary to the claims of Kundu \& Whitmore, we find that the observed metallicity
distributions are in close agreement with the predictions of such hierarchical formation models.
The mass spectrum of protogalactic fragments for the galaxies in our sample
has a power-law behavior, $n({\cal M}) \propto {\cal M}^{\alpha}$, with an
index of $\alpha \simeq -2$. This spectrum is indistinguishable from the mass spectrum of 
dark matter halos predicted by cold dark matter models for structure 
formation. We argue that these protogalactic fragments --- the likely sites of globular 
cluster formation in the early universe --- are the disrupted remains of the ``missing"
satellite galaxies predicted by cold dark matter models. Our findings suggest that
the solution to the missing satellite problem is through the suppression
of gas accretion in low-mass halos after reionization, or via self-interacting 
dark matter, and argue against models with suppressed small-scale power or 
warm dark matter.
\end{abstract}

\keywords{Cosmology: observations, Cosmology: dark matter, Galaxies: Elliptical and Lenticular, 
cD, Galaxies: Formation, Galaxies: Star Clusters}

\section{Introduction}

Cold dark matter (CDM) models offer an attractive theory for the formation of 
cosmological structure. These models, and particularly the $\Lambda$CDM variant, show impressive 
self-consistency with many of the most fundamental astrophysical observations: 
$i.e.,$ the expansion age of the universe, the primordial element abundance, the two-point correlation
function of cluster and galaxies, the local space density and redshift evolution of massive
clusters, and the large-scale streaming motions of galaxies in the local universe 
($e.g.,$ Bahcall et al. 1999). Nevertheless, it has become clear that this 
success on large scales ($i.e.,$ $R \gae$ 0.1-1 Mpc) does not extend to the 
level of galaxies (see, $e.g.,$ Sellwood \& Kosowsky 2001 for a review). Simply stated,
CDM models produce an overabundance of power on small scales when normalized to match the 
observed large-scale power: $e.g.,$ the models yield dark matter density profiles that 
are steeper than those observed (Flores \& Primack 1994), and
overpredict the number of dwarf galaxies (Kauffmann, White \& Guideroni 1993). We examine the
latter issue --- the so-called ``missing satellite problem" --- by using the globular
clusters (GCs) associated with a sample of nearby, early-type galaxies to derive
the mass distribution of accreted material in these systems.

Most recent attempts to explain the missing satellite problem have focussed on the properties 
of the dark matter itself. Competing models suggest that the dark matter is warm (Col\'in, 
Avila-Reese \& Valenzuela 2000), decaying (Cen 2001), self-interacting (Dav\'e et al. 2001), 
repulsive (Goodman 2000) or annihilating (Riotto \& Tkachev 2000). 
Alternatively, the standard CDM picture may be modified to include a broken scale invariance 
which suppresses small-scale power (Kamionkowski \& Liddle 2000). Yet another 
modification has been proposed by Bullock et al. (2000), who note that
a photo-ionizing background is expected to inhibit the accretion of gas onto low-mass halos after
the epoch of reionization. In this latter scenario, small-scale power is not suppressed, but
{\it hidden}: as a mixed population of disrupted and surviving dark-matter-dominated subhalos.

In what follows, we present a simple technique for measuring the mass distribution of disrupted 
subhalos in low-redshift galaxies. Our technique relies on the GC systems of these galaxies 
which, by virtue of their extreme age and apparently universal formation efficiency,
represent ideal tracers of the original population of low-mass subhalos or, in the
terminology used here, protogalactic fragments. 
The observational database consists of GC color 
and metallicity distributions for 28 early-type galaxies presented by Kundu \& Whitmore (2001). 
Although most cosmological evidence appears to favor the notion that the GC systems of 
large galaxies were assembled hierarchically ($e.g.,$ see Gnedin, Lahav \& Rees 2001 and
references therein), Kundu \& Whitmore (2001) claim that the GC metallicity distributions
of their program galaxies are incompatible with such scenarios. In this paper, we analyze the 
data of Kundu \& Whitmore (2001) using a quantitative numerical technique, described in a pair 
of earlier papers (C\^ot\'e, Marzke \& West 1998; C\^ot\'e et~al. 2001), which is designed to
recover the protogalactic mass spectrum of a galaxy. The method assumes that the mergers and accretions
by which cause galaxies to grow hierarchically induces little or no gas dissipation. The mass spectra
which we derive for the protogalactic fragments in each of these galaxies are indistinguishable 
from those predicted by CDM models for the population of low-mass dark matter 
halos, a finding which has obvious implications for the missing satellite problem.

\section{Globular Cluster Data}

Since GCs as a class are composed of coeval stars, the interpretation of their broadband colors 
is far more straightforward than is the case for galaxies. Moreover, for
composite stellar populations with mean ages of a few Gyr or more, broadband
colors are primarily sensitive to metallicity, not age (Bruzual \& Charlot 1993; Worthey 1994). 
Luckily, the great majority of the GCs in these systems appear to be older than this, according
to optical spectroscopy, color-magnitude diagrams, and luminosity function turnover magnitudes for 
GCs belonging to nearby dwarf elliptical, dwarf spheroidal, spiral and giant elliptical 
galaxies ($e.g.,$ Buonanno et al. 1998; Layden \& Sarajedini 1997; Harris, Poole \& 
Harris 1998; Cohen, Blakeslee \& Rhyzov 1998; Puzia et al. 1999; Lee \& Kim 2000; Beasley 
et al. 2000; Larsen et al. 2001). Indeed, evidence from various sources (see Gnedin et~al.
2001) points to minimum GC ages in the range $8 \lae t \lae 12$ Gyr,
suggesting that, in the currently fashionable $\Lambda$CDM models,  the bulk of GC formation 
in early-type galaxies had occurred {\it before} $1 \lae z \lae 4$. 
Thus, for the early-type galaxies considered here, we may be confident that the observed 
GC colors are reliable metallicity indicators.

Several recent studies have reported color/metallicity distributions for the
GC systems of early-type galaxies based on archival {\it HST} data ($e.g.,$
Gebhardt \& Kissler-Patig 1999; Kundu \& Whitmore 2001; Larsen et al. 2001). As noted by van den
Bergh (2001), the GC systems of these galaxies exhibit ``a bewildering variety
of characteristics". This is particularly true of their metallicity distributions which range 
from obviously multi-modal systems to those with a single (metal-rich or
metal-poor) peak. This remarkable variety has led some
investigators to suggest that no single recipe for galaxy formation is capable of explaining 
the full range in GC properties ($e.g.,$ Harris 2000). Indeed, there is clear supporting
evidence for most of the models that seek to explain the formation of GC systems and their host
galaxies: $e.g.,$ the formation of massive clusters in gas-rich mergers ($e.g.,$ Whitmore \&
Schweizer 1995) and the acquisition of GCs through dissipationless accretions ($e.g.,$ Ibata,
Gilmore \& Irwin 1994). The most immediate question is which --- if any --- of these processes 
has dominated the formation of the GC systems which we observe today.

In what follows, we model these diverse GC metallicity distributions using a simple prescription 
for hierarchical galaxy formation, described more fully in \S 3. In the interests of homogeneity, 
we limit ourselves to the GC color and metallicity distributions for 28 early-type galaxies presented 
in Kundu \& Whitmore (2001). Their galaxy sample is $\sim$ 60\% larger than that of Larsen et al. 
(2001), while the number of detected GCs per galaxy is $\sim$ 50\% larger than that of
Gebhardt \& Kissler-Patig (1999) for galaxies appearing in both studies. A single 
galaxy in the Kundu \& 
Whitmore sample, NGC~4486B, has been excluded from the analysis since the modeling of 
its GC metallicity distribution is complicated by the fact that a significant fraction of its
initial mass has probably been tidally stripped by NGC~4486 ($e.g.$, Faber 1973; Kormendy 
et~al. 1997).

Kundu \& Whitmore (2001) present VI photometry and (V-I) colors for GC candidates associated with
each of their program galaxies. The number of GCs, $N_{\rm GC}$, ranges between 445 for NGC~4649 and
41 for NGC~5845, with a mean of $\langle N_{\rm GC}\rangle$ = 163. Since our goal is to model
the detailed shape of the observed GC metallicity distributions, we restrict ourselves to
GCs with colors measured to a precision of $\sigma$(V-I) $=$ 0.2 mag or better --- roughly the separation
between the metal-rich and metal-poor peaks in the most luminous galaxies. With this selection
criterion,
the number of GCs per galaxy ranges from 404 in NGC~4649 to 41 in NGC~5845, with a mean value 
of $\langle N_{\rm GC}\rangle$ = 144. The adopted GC color-metallicity relation is
that of Kissler-Patig et~al. (1998),
$${\rm [Fe/H]} = -4.50 + 3.27{\rm (V-I)}  \eqno{(1)}$$
which is based on a combined sample of GCs in the Milky Way and NGC~1399 with measured
metallicities. Although there is significant scatter about this best-fit relation (see Figure~13 of
Kissler-Patig et~al. 1998), our conclusions are largely insensitive to the exact choice of 
color-metallicity relation.  The clusters used to establish this relation have metallicities in 
the range $-2.4 \lae$ [Fe/H] $\lae +0.3$ dex (0.64 $\lae$ V-I $\lae$ 1.47 mag), an interval 
which includes the vast majority of clusters considered here. 

To simulate the metallicity distribution of a GC system, we require an estimate
of the host galaxy's luminosity, and hence distance. In principal, it is possible to measure 
individual galaxy distances from the GC data themselves, using the observed turnover of the 
GC luminosity function (GCLF). Unfortunately, 
Kundu \& Whitmore (2001) were able to obtain reliable two-parameter GCLF fits ($i.e.,$ turnover 
magnitude and dispersion) for just 10 of their 28 galaxies. As a consequence, distances for 
their program galaxies were taken primarily from early versions of the 
surface brightness fluctuation (SBF) survey of Tonry and collaborators, with extinction 
estimates from Burstein \& Heiles (1984).
Recently, the final version of the SBF catalog --- based on an improved calibration --- 
has become available (Tonry et al. 2001), so we adopt the latest SBF distances for the sample 
galaxies. Likewise, improved extinctions for all 28 galaxies have been calculated from the 
Schlegel, Finkbeiner \& Davis (1998) DIRBE maps. Unfortunately, two of the sample galaxies
--- NGC~7626 and NGC~5982 --- are not included in Tonry et al. (2001) catalog.
For NGC~7626, we have combined the Schlegel et al. (1998) extinction with
the Tully-Fisher distance for the Pegasus group from Sakai et al. (2000) to find a 
true distance modulus of 33.73 mag. This is significantly larger
than the value of 33.01 adopted by Kundu \& Whitmore based on the fundamental plane
distance of Prugniel \& Simien (1996). For NGC~5982, no recent distance determination is
available, so we follow Kundu \& Whitmore in adopting the Prugniel \& Simien (1996)
distance for this galaxy, but corrected for our new extinction. Because of these revised 
distances and extinctions, the absolute magnitudes of the program galaxies 
differ from those reported in Kundu \& Whitmore. The mean magnitude 
difference is $\langle{\Delta}M_V\rangle$ = 0.18$\pm$0.08 mag (mean error),
in the sense that our magnitudes are brighter than those of Kundu \& Whitmore. The
largest discrepancies are found for NGC~5813, NGC~7626 and IC~1459, whose absolute magnitudes
have become brighter by ${\Delta}M_V \sim$ 0.9 mag.

Observational data for the sample galaxies are summarized in Table~\ref{tab1}. From left to right, 
the first six columns in this table report the galaxy name, integrated V-band magnitude from NED, 
extinction from Schlegel et al. (1998), adopted distance modulus, absolute visual magnitude, and total 
number of GCs having ${\sigma}$(V-I) $\le 0.2$ mag. The remaining columns are discussed in detail below.

\section{Numerical Method}

\subsection{Model Input}

We model the observed GC color and metallicity distributions for our program galaxies using
a numerical code that has been described in two previous papers. The first of these papers 
(C\^ot\'e et~al. 1998) outlined the strategy for simulating GC color and metallicity
distributions of galaxies that grow hierarchically via dissipationless mergers, and applied the
method to the observed GC metallicity distribution of the giant elliptical galaxy NGC~4472, one of the objects
in the current sample. A number of refinements to this original code were described in C\^ot\'e 
et~al. (2000), where the model was extended to the GC systems and halo field stars of spiral 
galaxies ($i.e.,$ the Milky Way and M31). Since a complete discussion of the approach may be 
found in these papers, we give only a brief outline here.

The method uses an empirical, Monte-Carlo approach to simulate the metallicity distributions of GCs
associated with the end-products of multiple dissipationless mergers and accretions, as expected
in hierarchical models in which the {\it bulk} of GC (and star) formation takes place in
distinct protogalactic fragments and occurs {\it prior} to the epoch of galaxy 
assembly.\altaffilmark{1}\altaffiltext{1}{Strictly speaking, the process is dissipationless
only in the sense that the formation of GCs --- clearly a dissipative process --- is not driven by
the mergers themselves, but rather, by the cooling and collapse of gas with each protogalactic
fragment.}
Using a Monte-Carlo approach, it is possible to simulate the GC metallicity distribution of
an individual galaxy by specifying three model ingredients: (1) an assumed form for 
the mass or luminosity function of the protogalactic fragments which agglomerate into
the final galaxy; (2) a relation between the mass or luminosity of the protogalactic fragment 
and the mean metallicity of its GC system; and (3) the number of GCs per unit protogalactic 
fragment mass or luminosity. Each of these ingredients is determined empirically from
observations of the GC systems of low-redshift galaxies. This empirical approach has the 
important practical advantage that it avoids the uncertainties involved in modeling directly the 
cycle of gas cooling, star formation and feedback in these halos (issues which, ultimately,
must be addressed in any complete theory of GC formation). As discussed in C\^ot\'e et~al. (1998),
the merging probability of protogalactic fragments is assumed to be independent of mass. 
As Figure~1 of Bullock et~al. (2001) shows for the specific case of the Milky Way, this 
assumption appears remarkably sound for protogalactic fragments with circular velocities 
of less than $v_c \simeq 60$ km~s$^{-1}$, corresponding to a mass of 
${\cal M} \sim$ a few times $10^9~{\cal M}_{\odot}$.

The luminosity function of protogalactic fragments is parameterized as a Schechter function,
$$n({\cal L}) \propto ({\cal L}/{\cal L}^*)^{\alpha} \exp{(-{\cal L}/{\cal L}^*)} \eqno{(2)}$$
where ${\cal L}^*$ is a characteristic luminosity and $\alpha$ is an exponent which governs the 
relative number of faint and bright systems (Schechter 1976). Note that the luminosity
in equation~(2) refers to the {\it present day} luminosity of surviving protogalactic 
fragments ($i.e.,$ isolated galaxies of low- and intermediate- luminosity; Larson 1988).
Over the luminosity range considered here, we assume a constant $V$-band mass-to-light ratio of 
${\Upsilon}_V = 5$ at $z = 0$, which is roughly appropriate for the low- and 
intermediate-luminosity galaxies used below to define the relation between mean 
GC metallicity and host galaxy luminosity (Mateo 1998). We take the low-luminosity cutoff in
equation~(2) to be ${\cal L}_V = 1.6\times10^7~{\cal L}_{V,{\odot}}$, or
${\cal M} = 8\times10^7~{\cal M}_{\odot}$. These limits correspond to the 
Fornax dSph galaxy (the faintest galaxy known to contain its own GC system) but
the precise choice has little effect on the calculated GC metallicity distributions for 
slopes shallower than $\alpha = -2$.  At the bright end, we adopt a constant value 
of ${\cal L}_V^* = 2\times10^{10}~{\cal L}_{V,{\odot}}$ for all
simulations (C\^ot\'e et~al. 2000), while $\alpha$ is a free parameter to be determined
on a galaxy-by-galaxy basis through a comparison of the observed and simulated GC
metallicity distributions. Note that, for masses less than ${\cal M} \sim 10^{11}~{\cal M}_{\odot}$,
the protogalactic mass function has a nearly power-law form,
$$n({\cal M}) \propto ({\cal M}/{\cal M}^*)^{\alpha} , \eqno{(3)}$$
which is also characteristic of the extended
Press-Schechter mass functions used in semi-analytic models.
Our decision to treat the change in the initial luminosity and mass
functions as a change in $\alpha$ alone is clearly an approximation, as
${\cal L}^*$ and $\cal M^*$ must have unavoidably increased in the presence of 
mergers and accretions. Although our simulations have no time resolution,
cosmological N-body simulations
demonstrate that the change in $\cal M^*$ should have only a modest effect on
the final GC metallicity distribution: Figure~12 of Pearce et~al. (2001)
suggests the that change in the high-end cutoff of the mass function between $z = 3$ and $z = 0.2$ 
is about a factor of four, corresponding to a maximum shift in $\langle$V-I$\rangle$ color of 
0.08~mag for the highest mass protogalactic fragments.

The relation between GC metallicity and luminosity/mass of the host protogalactic fragment 
has been discussed extensively in C\^ot\'e et al. (2000). Figure~\ref{fig1} shows the mean GC
metallicity plotted against the luminosity (or mass) of nine dwarf elliptical (dE) and 
dwarf spheroidal (dSph) galaxies (circles), with data taken directly from that paper. Note
that GC metallicities for these galaxies have been converted into (V-I) colors according
to equation~(1).
Triangles indicate the metal-rich GC systems of two well-studied spiral galaxies: the Milky 
Way and M31. In both cases, the luminosities and masses refer to their {\it bulges}; 
that is to say, we assign the metal-rich GCs in these galaxies to their bulge components,
and not their disks (Minniti 1995; C\^ot\'e 1999).  Our estimate for the relation between the mean GC 
color and the absolute magnitude of the host protogalactic fragment is
$$\langle{\rm V-I}\rangle = 0.04 - 0.055M_V, \eqno{(4)}$$
which is indicated by the dashed line in Figure~\ref{fig1}. This ``zero-age" relation
gives the dependence of mean GC color on the luminosity, or mass, of the progenitor 
protogalactic fragment, {\it as it would appear today}, if the subsequent evolution has been 
mostly passive. In the simulations, the mean color/metallicity of GCs belonging to
each protogalactic fragment is determined using equation~(4), with an assumed 1$\sigma$
scatter of 0.03~mag (0.1 dex). The {\it internal} dispersion in color for each protogalactic
fragment is also approximated as a Gaussian, with dispersion 0.09~mag (0.29 dex)
based on the findings of C\^ot\'e et~al. (2000).

Finally, we adopt a constant GC specific frequency for all protogalactic fragments ---
$S_N \equiv 5\pm1$ at the present time --- and neglect possible variations in $S_N$
at the low- and high-mass ends of the galaxy distribution (see \S 6). 
In the notation of McLaughlin (1999), this specific frequency corresponds to a
GC formation efficiency of
$$\epsilon_{\rm GC} = 0.29\pm0.06\% \eqno{(5)}$$
by mass. This is consistent with the value of $\epsilon_{\rm GC} = 0.26\pm0.05\%$ advocated
by McLaughlin (1999) based on observations of about one hundred early-type galaxies. The 20\%
uncertainty in our adopted values of $S_N$ and $\epsilon_{\rm cl}$ reflects the observed 
scatter about their means, and has been included explicitly in our Monte-Carlo simulations.

\subsection{An Example}

Before proceeding, we present a descriptive example of how the GC system of a galaxy 
that grows hierarchically through dissipationless mergers will evolve in the color-magnitude 
plane. Consider a protogalactic fragment, A, with absolute magnitude $M_V = -20$~mag and
mass ${\cal M} = 4\times10^{10}~{\cal M}_{\odot}$ which falls on the zero-age 
color-magnitude relation ($i.e.,$ the dashed line in Figure~\ref{fig1}). According to 
equation~(4), the unimodal GC system associated with this protogalactic fragment will 
have $\langle$V-I$\rangle$ = 1.14~mag. First, suppose that this fragment merges
with an identical fragment, producing an end-product, C1, with $M_V = 20.75$~mag
and a GC system having a single peak at $\langle$V-I$\rangle$ = 1.14~mag.
Alternatively, fragment A could accrete a population of 100 protogalactic fragments, B, 
each of which has $M_V = -15$~mag, and $\langle$V-I$\rangle$ = 0.865~mag.  But, by virtue of 
equation (5) and our assumed constant mass-to-light ratio, fragment A contributes the same 
number of GCs as the 100 smaller fragments combined. Thus, the GC systems of the final 
galaxy, C2, will have two distinct components (containing equal numbers of metal-rich 
and metal-poor GCs), and a mean color of $\langle$V-I$\rangle \simeq$ 1.00~mag.
A third possible end-product, C3, is also shown in Figure~\ref{fig1}. In this
case, the galaxy was formed entirely through the agglomeration of (200) fragments
of type B, so that the final GC system is unimodal, but with a metal-poor peak at 
$\langle$V-I$\rangle$ = 0.865~mag.

The point of this simple exercise to emphasize that, once the ``zero-age" color-metallicity
is fixed, it is possible --- through a suitable adjustment of the mass spectrum of the 
protogalactic fragments --- to reproduce the GC systems of {\it any} galaxy whose GC system
falls within the dotted region shown in Figure~\ref{fig1}. This simple recipe for
galaxy formation combines dissipationless mergers, which serve to move the galaxy to the right of the
initial color-magnitude relation, with a dissipative, monolithic collapse scenario for the constituent
protogalactic fragments. Our goal is to use the observed GC metallicity distribution for a 
specific galaxy and solve an inverse problem to determine the mass spectrum of protogalactic 
fragments which best matches the observed metallicity distribution.

\subsection{Application}

With equations~(2) and (4) specified, it is a trivial matter to calculate analytically the 
metallicity distribution of a GC system which arises from dissipationless mergers
and accretions. In practice, such an approach is
thwarted by the highly stochastic nature of hierarchical galaxy formation ($e.g.,$ Lacey \& 
Cole 1993), particularly at the high-mass end of the protogalactic mass spectrum.
Fortunately, this stochasticity can be used to reconstruct merger histories for
individual galaxies since the fossil record of the largest protogalactic fragments is
ussually preserved in the final GC metallicity distribution. 

To do so, we generate, for each galaxy in our sample, 50 simulated GC systems at each point 
over a 10$\times$10 grid in ($\zeta$, $\alpha$) where $\zeta$ is the ratio
between the mass of the largest protogalactic fragment and the final mass of the galaxy,
$$\zeta= {\cal M}^1/{\cal M}_{\it f}, \eqno{(5)}$$
and $\alpha$ is the slope of the protogalactic mass function. The ($\zeta$, $\alpha$) plane is
sampled at $\zeta = 0.1,0.2,...1.0$ and and $\alpha = -1.1, -1.2, ... -2.0$.
For each of the 5000 simulations per galaxy, we extract at random from the full GC system the same
number of GCs as contained in the Kundu \& Whitmore (2001) catalog,
add the appropriate amount of measurement error to the simulated (V-I) colors, and bin
the actual and simulated data in an identical manner (see \S 4.1). A goodness-of-fit statistic,
$${\chi}^2 = {1 \over N_{\rm bin} - 1} {\sum_{i=1}^{N_{\rm bin}}} {({N_{{\rm{obs,}}i} - N_{{\rm{sim,}}i}})^2 \over (N_{{\rm{obs,}}i} + N_{{\rm{sim,}}i}) } \eqno{(6)}$$
is then calculated for each simulation, along with a mean $\chi^2$ for each point on the 10$\times$10 
grid. The number of bins, $N_{\rm bin}$, is chosen on a galaxy-by-galaxy basis as described in
\S 4.1.  As an illustration of the method, Figure~\ref{fig2} shows a contour plot over this grid 
for one the galaxies in our sample, NGC~4472. From these simulations, we deduce best-fit values
of $\zeta = 0.30\pm0.05$ and $\alpha = -1.85\pm0.10$ (1$\sigma$ uncertainties). 
The cross in Figure~\ref{fig2} shows the location in the ($\zeta$,~$\alpha$) plane of
the simulation which produced the best match to the observed GC metallicity distribution.
These results are summarized in the final three columns of Table~\ref{tab1} which record
the $\chi^2$ for the best-fit simulation, along with the corresponding values of 
$\alpha$ and $\zeta$ and their 1$\sigma$ uncertainties.

The mean simulated color of the GC systems in the sample is $\langle$V-I$\rangle_{\rm sim} = 1.00$, 
while the mean difference between the observed and simulated GC color is 
$\langle$V-I$\rangle_{\rm obs}$-$\langle$V-I$\rangle_{\rm sim} = 0.00$~mag, with a standard 
deviation of 0.01~mag. We now turn our attention to a comparison of the detailed shapes
of the observed and simulated GC color and metallicity distributions.

\section{Results}

\subsection{Metallicity Distributions}

In this section, we compare the observed GC color and metallicity distributions 
with the best-fit models for each of our program galaxies. We follow the usual
practice of presenting the results as histograms: $i.e.,$ as parametric
representations of the underlying distributions. An optimal histogram bin 
width, $h$, is calculated on a case-by-case basis according to the GC
sample size and level of non-Gaussianity for each metallicity distribution. 
Specifically, we use the method of Scott (1992) and set the bin width at
$$h \simeq {2.7{\sigma}{\eta} \over N_{\rm GC}^{1/3}} \eqno{(6)}$$
where ${\sigma}$ is the standard deviation of the distribution and
${\eta}$ is a factor in the range $0.4 \lae {\eta} \lae 1$ related
to the distribution's skewness. For our program objects, the bin widths range 
between $h \simeq 0.055$~mag for NGC~4649, and $h \simeq 0.138$ for NGC~4550;
these values of $h$ are comparable to the mean errors, $\sigma$(V-I) = 0.08~mag,
of the measured GC colors. Note that this
procedure differs from that of Kundu \& Whitmore (2001), who opted to fix the bin 
width at a constant value of $h$ = 0.03~mag, although $N_{\rm GC}$
varies by more than an order of magnitude among their program galaxies. 

Figures~\ref{fig3}-\ref{fig9} show the metallicity distributions for GCs belonging to the
28 galaxies listed in Table~\ref{tab1}. The observed distributions are indicated by the
filled circles, while the errorbars show the Poisson uncertainties in each bin. For
each galaxy, the best-fit model is indicated by the dashed curve, which has been binned
in the same manner as the actual data. 
The excellent agreement between the observed and simulated distributions is noteworthy in
light of the fact that, for the simple model considered here, the effects of just two
free parameters, $\zeta$ and $\alpha$, are explored. A more comprehensive exploration of
parameter space might include 
possible mass-dependent variations in the GCs formation
efficiency, the dynamical evolution of the GC systems, GC formation 
induced by the merger and accretion process, and the form of the intrinsic metallicity
distribution in the protogalactic fragments. The first two processes were
included in the Milky Way formation model discussed in C\^ot\'e et al. (2000), but have
been omitted here for the sake of simplicity. We note, however, that both the weak dependence 
of GC specific frequency on host galaxy luminosity ($e.g.,$ Durrell et~al. 1996; Miller et~al. 
1998; McLaughlin 1999) and the preferential disruption of the more
centrally concentrated metal-rich GCs in these galaxies, will serve to {\it increase}
the ratio of metal-poor to metal-rich GCs. Including these effects in the simulations 
would give additional freedom to match the relative numbers of metal-rich and metal-poor 
GCs in these galaxies. To first order, these processes affect the peak heights for the 
subsystems, and not their mean metallicities.

Figure~\ref{fig10} presents the same color-magnitude plane for GC systems that was shown 
in Figure~\ref{fig1}. However, we now include the GC systems of the 28 galaxies listed in 
Table~\ref{tab1}. Kundu and Whitmore (2001) classified each of these GC systems as either 
``certainly" bimodal, ``likely" bimodal or unimodal. We adopt their classifications verbatim. 
For the 16 galaxies classified as certainly or likely bimodal, the peak color
for the metal-rich and metal-poor GC systems (according to KMM) are indicated by the
filled squares and circles, respectively.\altaffilmark{2}\altaffiltext{2}{The GC systems of
NGC~3377 and NGC~584 are classified by Kundu \& Whitmore (2001) as ``likely" to be bimodal 
in their Table~\ref{tab1}, but plotted as ``certain" in their Figure 5. We have assumed
that the GC systems of both galaxies show statistically significant evidence of bimodality.} 
The filled triangles indicate the mean color and metallicity of the 12 unimodal GC systems.
The open symbols show the median colors and metallicities of the GC systems in 
each of these galaxies based on 100 simulations using the best-fit values of 
$\zeta$ and $\alpha$ listed in Table~\ref{tab1}. Simulations for the unimodal systems are indicated
by the open triangles, while the open squares and open circles show the median colors for
the metal-rich and metal-poor components in the bimodal systems (determined from the
best-fitting double Gaussians). To illustrate the typical
{\it range} in simulated GC color and metallicity, an errorbar is included in the lower right
corner of this figure, showing the 1$\sigma$ dispersion in the colors and metallicities 
based on the simulations.

The simulations reproduce the observed GC colors and metallicities of not just
the unimodal systems, but also of the metal-rich and metal-poor components in the bimodal 
GC systems. Likewise, the simulations reproduce the observed trend for the metal-poor peak in
the bimodal GC systems to depend only weakly on host galaxy luminosity (Forbes,
Brodie \& Grillmair 1997; C\^ot\'e et al. 1998; Larsen et~al. 2001) and, by definition, 
the clear
correlation between 
metal-rich peak and host galaxy luminosity. This latter trend can be traced to the
fact that the first-ranked protogalactic fragments: (1) themselves originate on the ``zero-age" 
color-magnitude relation; and (2) contribute a significant fraction of the galaxy's final mass. 
Indeed, for each galaxy classified by Kundu \& Whitmore as bimodal, we show the 
initial location in the color-magnitude plane of the largest 
protogalactic fragment, determined from the best-fit simulation for each galaxy (open stars). 
Thin arrows connect these initial positions to the
locations of the metal-rich peak measured directly from the observed GC metallicity
distribution. Note that, in a few cases, measurement errors and observational scatter 
(indicated by the errorbar) cause the initial location in the color-magnitude plane to lie 
{\it below} the final location of the metal-rich peak.
The filled pentagons show the GC systems of the same protogalactic 
fragments ($i.e.,$ dwarfs and spiral bulges) used to define the ``zero-age" color-magnitude
relation in Figure~\ref{fig1}, reproduced as the dashed line in Figure~\ref{fig10}. 
We conclude that the simple model considered here provides an adequate match to both the 
observed GC metallicity distributions {\it and} the final location of the GC systems 
in the color-magnitude plane.

\subsection{Comparisons with Previous Results}

On the basis of the same GC data analyzed here, Kundu \& Whitmore (2001) claim that hierarchical 
models for galaxy formation are unviable.  As Figures~\ref{fig3}-\ref{fig10} demonstrate, this 
conclusion is not supported by their data.

Kundu \& Whitmore (2001) made no attempt to simulate the GC color distributions expected 
in hierarchical formation scenarios, despite the fact that the implementation of such
algorithms has been described in detail previously ($e.g.,$ C\^ot\'e et al. 2000; 
Harris et~al. 2000). Instead, they advocate a qualitative scenario in which 
GC formation occurs in multiple star formation episodes driven by mergers of gas-rich disk 
galaxies, although such models are curently not amenable to observational tests.
They further note that some of the low-luminosity 
galaxies in their sample ($e.g.,$ NGC~1439, NGC~3377 and NGC~4660) are likely to have bimodal
GC metallicity distributions and claim that this observation poses ``a serious problem"
for the hierarchical model. This claim is surprising
since C\^ot\'e et al. (2001) showed that hierarchical models were able to reproduce the
bimodal GC metallicity distribution of the Galactic spheroid, which is several times fainter than
NGC~1439, NGC~1427 and NGC~3377. Moreover, all of the low-luminosity galaxies having bimodal
GC metallicity distributions fall within the region in Figure~\ref{fig1} that is expected to be
occupied by galaxies which form hierarchically (\S 3.2). As Figures~\ref{fig3}-\ref{fig9} attest, 
the observed diversity in the GC metallicity distributions {\it and} the bimodal nature of some 
low-luminosity ellipticals could arise naturally within the context of a simple hierarchical
model.

Kundu \& Whitmore (2001) also state that the success of the hierarchical model
``is critically dependent on the second-order nature of [the metallicity-luminosity relation for] 
the systems being cannibalized". However, this claim too is incorrect since the model distributions
presented in Figures~\ref{fig3}-\ref{fig9} are derived from a {\it linear} metallicity-luminosity 
relation (see \S 3.1), not the second-order relation adopted by C\^ot\'e et al. (1998) on
these basis of preliminary data. While it is clearly important to further refine the
metallicity luminosity relation for the GC systems of low- and intermediate-luminosity
galaxies, the model described here is quite general in nature, and our basic conclusions are 
unlikely to be changed by minor modifications to the metallicity-luminosity relation.

\subsection{Protogalactic Mass Spectra}

Having shown that hierarchical growth through dissipationless mergers and accretions offers a 
viable explanation for the diverse metallicity distributions of GC systems in early-type galaxies, we now
examine properties of protogalactic fragments from which the galaxies may have been assembled.
Figures~\ref{fig11}-\ref{fig17} show protogalactic luminosity and mass functions for each galaxy,
based on the simulation which matches the observed GC metallicity distribution.
The shaded histogram in each panel shows the best estimate for the distribution of protogalactic 
fragments, while the dashed curve indicates the parent Schechter function from which the 
simulation was drawn. For all galaxies, the corresponding GC metallicity distributions are
indicated by the dashed curves in Figures~\ref{fig3}-\ref{fig9}.

A striking feature of the protogalactic luminosity and mass functions shown in 
Figures~\ref{fig11}-\ref{fig17} is their steep slope. From Table~\ref{tab1}, the 
average power-law exponent for the protogalactic luminosity and mass functions 
is $\langle\alpha\rangle = -1.88\pm0.03$ (mean error), with a standard deviation 
of $\sigma$($\alpha$) = 0.13.  This value is consistent with previous findings for 
NGC~4472 ($\alpha \sim -1.8$; C\^ot\'e et~al. 1998), the Milky Way ($\alpha \sim -2.0$; 
C\^ot\'e et al. 2000) and
M31 ($\alpha \sim -1.8$; C\^ot\'e et al. 2000). In the context of hierarchical formation
models, it is the apparent universality of this exponent which, when coupled with the fixed 
low-mass cutoff of the protogalactic luminosity function, conspires to produce the population 
of metal-poor GCs with [Fe/H]~$\sim -1.2~$dex that are present in virtually every large 
galaxy studied to date ($e.g.,$ Ashman \& Bird 1993; Brodie, Larsen \& Kissler-Patig 2000).
As the same time, the observed diversity of GC metallicity distributions arises from the 
stochastic fluctuations about this ``universal" protogalactic mass spectrum,
particularly at the high-mass end. By virtue of their large mass, these fragments 
contribute the bulk of the metal-rich GCs associated with the final galaxy.

From Table~\ref{tab1}, we see that there is a fairly wide range in the properties of the
most massive protogalactic fragments. On average, the most massive fragment contributes 
a fraction $\langle\zeta\rangle  = 0.25\pm0.03$ (mean error) of the galaxy's final mass, with a
standard deviation of $\sigma$($\zeta$) = 0.14. For some galaxies, such as NGC 4406, the
contribution of the largest fragment amounts to no more $\sim$ 10\% of the final mass,
while in NGC 4458, as much as 70\% of the final mass may have arisen in the dissipational
collapse of the largest protogalactic fragment. For NGC~4472, we find a best-fit value of 
$\zeta = 0.30\pm0.05$ (mean error), which compares favorably with the value 
of $\zeta \sim 0.35$ estimated by C\^ot\'e et al. (1998) using ground-based observations of its 
GC system carried out in a different filter system ($i.e.,$ $CT_1$ photometry from 
Geisler, Lee \& Kim 1996), and using an alternative color-magnitude relation.

Recent studies have concluded that approximately 50\% of bright, early-type
galaxies show evidence for bimodal color distributions (Gebhardt \& Kissler-Patig 1999;
Larsen et~al. 2001; Kundu \& Whitmore 2001), so it is of interest to ask how often the 
simulations yield a metallicity distribution that is bimodal at a statistically 
significant level. Unfortunately, the answer to this question depends rather sensitively
on the properties of each dataset used to assess the level of bimodality. To understand why 
this is so, consider Figure~\ref{fig18}, which shows the metallicity distribution for a 
representative program galaxy, NGC 5322. For the purposes of illustration, the solid
curve shows the (normalized) metallicity distribution corresponding to the protogalactic mass 
function plotted in the upper panel of Figure~\ref{fig12}, where the intrinsic dispersion
in metallicity for each protogalactic fragment is taken to be 0.01 dex. As discussed in 
\S 3.1 and in C\^ot\'e et al. (2000), actual fragments are likely to have had intrinsic
dispersions much larger than this, but this extreme example serves to illustrate
the complex nature of the underlying metallicity distribution, and demonstrates 
the dangers of interpreting multi-modal metallicity distributions as {\it prima facie} evidence 
for multiple bursts of GC formation. The remaining
curves show this same (renormalized) distribution after smoothing with Gaussians having dispersions of
0.1, 0.2, 0.3 and 0.4 dex (intended to represent the combined effects of an intrinsic 
metallicity dispersion {\it and} observational errors).  This last distribution is, like 
the observed distribution, ostensibly unimodal in nature (Kundu \& Whitmore 2001), but 
this is mainly a consequence of the intrinsic
metallicity dispersion ($\sim$~0.3 dex) for the protogalactic fragments, and the $\sim$~0.3 dex
dispersion associated with metallicity measurements derived from (V-I) colors having
a precision of roughly 0.1 mag. Thus, establishing the frequency of bimodality depends
rather sensitively on the precision of metallicity measurements (and on other factors
including the total luminosity of the final galaxy and its detailed merger tree).

\section{Implications for the Missing Satellite Problem}

As mentioned in \S 1, one of the most troubling problems facing CDM models on small scales
is the so-called ``missing satellite" problem. Simply stated, the number of low-mass
dark halos predicted by CDM models exceeds the observed number of dwarf galaxies by factors
of tens, or even hundreds ($e.g.,$ Kauffmann, White \& Guiderdoni et~al. 1993; Klypin et~al. 
1999; Moore et al. 1999). Is it possible to explain the underabundance of dwarf galaxies in 
the local universe, and at the same time retain the basic framework of CDM models?

Figure~\ref{fig19} shows the ensemble spectrum of protogalactic fragments in
our sample: $i.e.,$ summed over all 28 galaxies (filled circles). As with the individual
galaxies, the composite mass function is found to follow a power-law behavior, 
$n({\cal M}) \propto {\cal M}^{\alpha}$, with exponent 
$\alpha \simeq -2$. While this mass function is much steeper than
the local galaxy luminosity function ($i.e.$, Pritchet \& van den Bergh 1999;
Folkes et~al. 1999; Blanton et~al. 2001), it is nevertheless in
remarkable agreement with the mass spectrum of dark matter halos predicted by
CDM models of structure formation; the dotted line in
Figure~\ref{fig19} shows the analytical approximation, $n({\cal M}) \propto {\cal M}^{-2}$,
for the mass spectrum of dark matter halos in the N-body simulations of Moore et~al. (1999). 
{\it In short, we find evidence for a disrupted population of dwarf-sized protogalactic 
fragments in each of our program objects, whose numbers and masses closely
resemble those of the predicted population of ``missing" satellite galaxies.}

This finding may have important consequences for the various models which have been
advanced as possible solutions to the missing satellite problem.
Such a disrupted population is, in fact, predicted by models
which account for the missing satellites by suppressed gas accretion
in low-mass halos after the epoch of reionization due the presence of a strong photo-ionizing
background ($e.g.,$ Rees 1986; Thoul \& Weinberg 1996; Bullock, Kravtsov \& Weinberg 2000).
A disrupted component of this sort is also expected in models which seek to explain the
missing satellite problem through self-interacting CDM ($e.g.,$ Spergel \&
Steinhardt 2000; Dav\'e et al. 2001). By contrast, a disrupted component is not expected
in models which explain the underabundance of dwarf galaxies through an {\it ab initio} 
reduction in small scale power 
($e.g.,$ Kamionkowski \& Liddle 2000) or by suppressing the formation of low-mass halos via 
warm dark matter ($e.g.,$ Col\'{\i}n, Avila-Reese \& Valenzuela 2000; Sommer-Larsen \& 
Dolgov 2001). Thus, these models appear to be 
inconsistent with the observed metallicity distributions of GCs in galaxies at $z \sim 0$ since 
the same low-mass halos whose formation is suppressed in these scenarios are actually
{\it required} to explain the GC metallicity distributions of luminous galaxies in 
hierarchical cosmologies.

\section{Summary and Conclusions}

We have described a technique for simulating the metallicity distributions of GCs belonging
to galaxies that grow hierarchically through dissipationless mergers and accretions. 
Applying the model to the GC systems of 28 early-type galaxies for which high quality {\it HST} 
data are available demonstrates such a mechanism is capable of explaining quantitatively
the observed diversity in extragalactic GC systems. Thus, claims that
the metallicity distributions of GCs in early-type galaxies are
inconsistent with a hierarchical origin (Kundu \& Whitmore 2001) are 
unsubstantiated. 

Protogalactic mass spectra for the sample of early-type galaxies considered here reveal
a striking similarity: they appear to follow a power-law distribution in mass, with 
$n({\cal M})~\propto~{\cal M}^{-2}$. This mass spectrum is indistinguishable from the
mass spectrum of dark matter halos predicted by CDM models, and thus has an important
implication for the missing satellite problem (Klypin et~al. 1999; Moore et~al. 1999).
Specifically, we find evidence from the GC systems of the galaxies examined
here for a disrupted population of dwarf-sized protogalactic
fragments whose numbers and masses closely
resemble those of the predicted population of missing satellite galaxies. 

If the basic scenario outlined here is correct ($i.e.,$ the early formation of GCs in low- 
and intermediate-mass halos after a rapid period of mass-dependent chemical enrichment,
followed by the agglomeration of these halos into the main body of the galaxy), then there
may be some interesting observational consequences of this picture.
For instance, if most of the low-mass dark matter halos predicted by CDM models have indeed
been rendered undetectable as a result of disruption and the suppression of gas accretion
after reionization, as suggested by Bullock et~al. (2000), then some fraction of the
original halo population may survive as ``dark galaxies" (see, $e.g.$, Klypin et~al.
1999 and references therein). Precise estimates for the redshift of GC formation 
are not yet available, but current evidence seems to favor formation at $z \gae 7$ 
(see Gnedin et~al. 2001), close to the expected edge of the epoch of reionization 
(Djorgovski et~al. 2001; Becker et~al. 2001; Barkana 2001). This raises the 
possibility that some of these putative ``dark galaxies", which have experienced
little or no star formation, might nevertheless have managed to form GCs.
Interestingly, there is some evidence that the number of
GCs per unit host galaxy luminosity increases toward fainter systems, at least
among dE,N galaxies (Durrell et~al. 1996; Miller et~al. 1998). Searches of
extragalactic GC systems for compact, dynamically-bound
groupings of GCs --- presumably metal-poor in nature and
not associated with an identifiable satellite galaxy --- might prove interesting.

\acknowledgments

We thank Arunav Kundu for providing the GC photometry used in the analysis, and an anonymous
referee for helpful comments.  MJW was supported by NSF grant AST00-71149.
This research has made use of the NASA/IPAC Extragalactic Database (NED) which is operated by the 
Jet Propulsion Laboratory, California Institute of Technology, under contract with the National 
Aeronautics and Space Administration.


\begin{deluxetable}{lrcccrcll}
\scriptsize
\tablecolumns{9}
\tablewidth{0pc}
\tablecaption{Protogalactic Mass Functions of Early-Type Galaxies Deduced From Their Globular Cluster Systems\label{tab1}}
\tablehead{
\colhead{Galaxy} &
\colhead{V$_{\rm T}$} &
\colhead{E(V-I)} &
\colhead{$5\log{d} -5$} &
\colhead{$M_V$} &
\colhead{$N_{\rm GC}$} &
\colhead{$\chi^2$} &
\colhead{$\alpha$} &
\colhead{$\zeta$} \\
\colhead{} &
\colhead{(mag)} &
\colhead{(mag)} &
\colhead{(mag)} &
\colhead{(mag)} &
\colhead{} &
\colhead{} &
\colhead{} &
\colhead{} 
}
\startdata
NGC~4472 &  8.41 & 0.030 &31.06& -22.72 & 338 &1.61&$-1.85\pm0.10$ & $0.30\pm0.05$ \\
NGC~4649 &  8.84 & 0.037 &31.13& -22.38 & 404 &0.65&$-1.90\pm0.10$ & $0.35\pm0.05$ \\
NGC~4406 &  8.90 & 0.039 &31.17& -22.37 & 181 &0.78&$-2.00^{+0.15}_{-0.05}$ & $0.10\pm0.05$ \\
NGC~4365 &  9.56 & 0.029 &31.55& -22.06 & 280 &0.67&$-1.65\pm0.15$ & $0.10^{+0.15}_{-0.05}$ \\
NGC~5322 & 10.23 & 0.016 &32.47& -22.29 & 118 &0.39&$-1.95^{+0.15}_{-0.05}$ & $0.30\pm0.1$ \\
NGC~4494 &  9.83 & 0.029 &31.16& -21.40 & 123 &0.46&$-2.00\pm0.05$ & $0.15^{+0.15}_{-0.10}$ \\
NGC~7626\tablenotemark{a} & 11.11 & 0.099 &33.73& -22.86 & 111 &0.44&$-1.60^{+0.15}_{-0.10}$ & $0.10\pm0.05$ \\
NGC~5982\tablenotemark{b} & 11.13 & 0.022 &32.87& -21.75 &  61 &0.39&$-1.95^{+0.25}_{-0.10}$ & $0.40\pm0.15$ \\
IC~1459  &  9.99 & 0.022 &32.38& -22.39 & 172 &0.86&$-1.95^{+0.05}_{-0.10}$ & $0.30^{+0.10}_{-0.05}$ \\
NGC~3610 & 10.84 & 0.014 &31.68& -20.84 & 107 &0.62&$-1.95^{+0.05}_{-0.10}$ & $0.45^{+0.10}_{-0.05}$ \\
NGC~~584 & 10.48 & 0.057 &31.66& -21.18 & 112 &0.69&$-1.90^{+0.20}_{-0.10}$ & $0.35^{+0.10}_{-0.15}$ \\
NGC~4621 &  9.63 & 0.024 &31.42& -21.79 & 165 &0.42&$-1.65^{+0.05}_{-0.10}$ & $0.25\pm0.05$ \\
NGC~4552 &  9.75 & 0.056 &31.07& -21.32 & 196 &1.11&$-2.00^{+0.20}_{-0.05}$ & $0.10^{+0.45}_{-0.05}$ \\
NGC~5813 & 10.46 & 0.076 &32.73& -22.27 & 185 &0.96&$-2.00\pm0.05$ & $0.10\pm0.05$ \\
NGC~4589 & 10.73 & 0.038 &31.80& -21.07 & 134 &1.03&$-1.95^{+0.15}_{-0.10}$ & $0.25\pm0.05$ \\
NGC~4278 & 10.16 & 0.038 &31.12& -20.96 & 244 &1.32&$-2.00\pm0.05$ & $0.30^{+0.05}_{-0.10}$ \\
NGC~4473 & 10.20 & 0.038 &31.07& -20.87 & 145 &0.44&$-2.00^{+0.10}_{-0.05}$ & $0.30\pm0.10$ \\
NGC~3379 &  9.28 & 0.033 &31.20& -20.92 &  63 &0.44&$-1.90^{+0.15}_{-0.10}$ & $0.35\pm0.10$ \\
NGC~~821 & 10.68 & 0.143 &32.26& -21.58 & 101 &1.19&$-2.00^{+0.10}_{-0.05}$ & $0.10^{+0.10}_{-0.05}$ \\
NGC~3608 & 10.76 & 0.029 &31.87& -21.11 &  95 &1.29&$-1.95^{+0.20}_{-0.10}$ & $0.10\pm0.05$ \\
NGC~4291 & 11.47 & 0.050 &32.21& -20.74 & 138 &0.84&$-1.95^{+0.15}_{-0.10}$ & $0.10\pm0.05$ \\
NGC~1439 & 11.39 & 0.041 &32.23& -20.84 &  85 &1.08&$-1.80^{+0.40}_{-0.20}$ & $0.10^{+0.10}_{-0.05}$ \\
NGC~1427 & 10.86 & 0.018 &31.90& -21.04 & 146 &0.54&$-2.00^{+0.20}_{-0.05}$ & $0.20^{+0.10}_{-0.15}$ \\
NGC~3377 & 10.38 & 0.046 &30.36& -19.98 & 106 &0.35&$-1.80^{+0.45}_{-0.15}$ & $0.40^{+0.10}_{-0.05}$ \\
NGC~4550 & 11.68 & 0.057 &31.13& -19.45 &  47 &0.83&$-1.75^{+0.50}_{-0.30}$ & $0.20\pm0.15$ \\
NGC~5845 & 12.48 & 0.073 &32.24& -19.76 &  41 &0.35&$-1.75^{+0.35}_{-0.30}$ & $0.35^{+0.25}_{-0.15}$ \\
NGC~4660 & 11.24 & 0.048 &30.64& -19.40 &  98 &0.53&$-1.65\pm0.15$ & $0.25^{+0.10}_{-0.05}$ \\
NGC~4458 & 12.07 & 0.033 &31.26& -19.19 &  43 &0.44&$-1.75\pm0.25$ & $0.70^{+0.10}_{-0.05}$ \\
& & & & & \multicolumn{2}{c}{} \\
         &       &       &       &   &     & Mean = & $-1.88\pm0.03$ & $0.25\pm0.03$ \\
\enddata
\tablenotetext{a}{Tully Fisher distance for the Pegasus group from Sakai et al. (2000), ApJ, 529, 698.}
\tablenotetext{b}{Fundamental Plane distance from Prugniel \& Simien (1996), A\&A, 309, 749}
\end{deluxetable}

%
%
 
\clearpage
 
\plotone{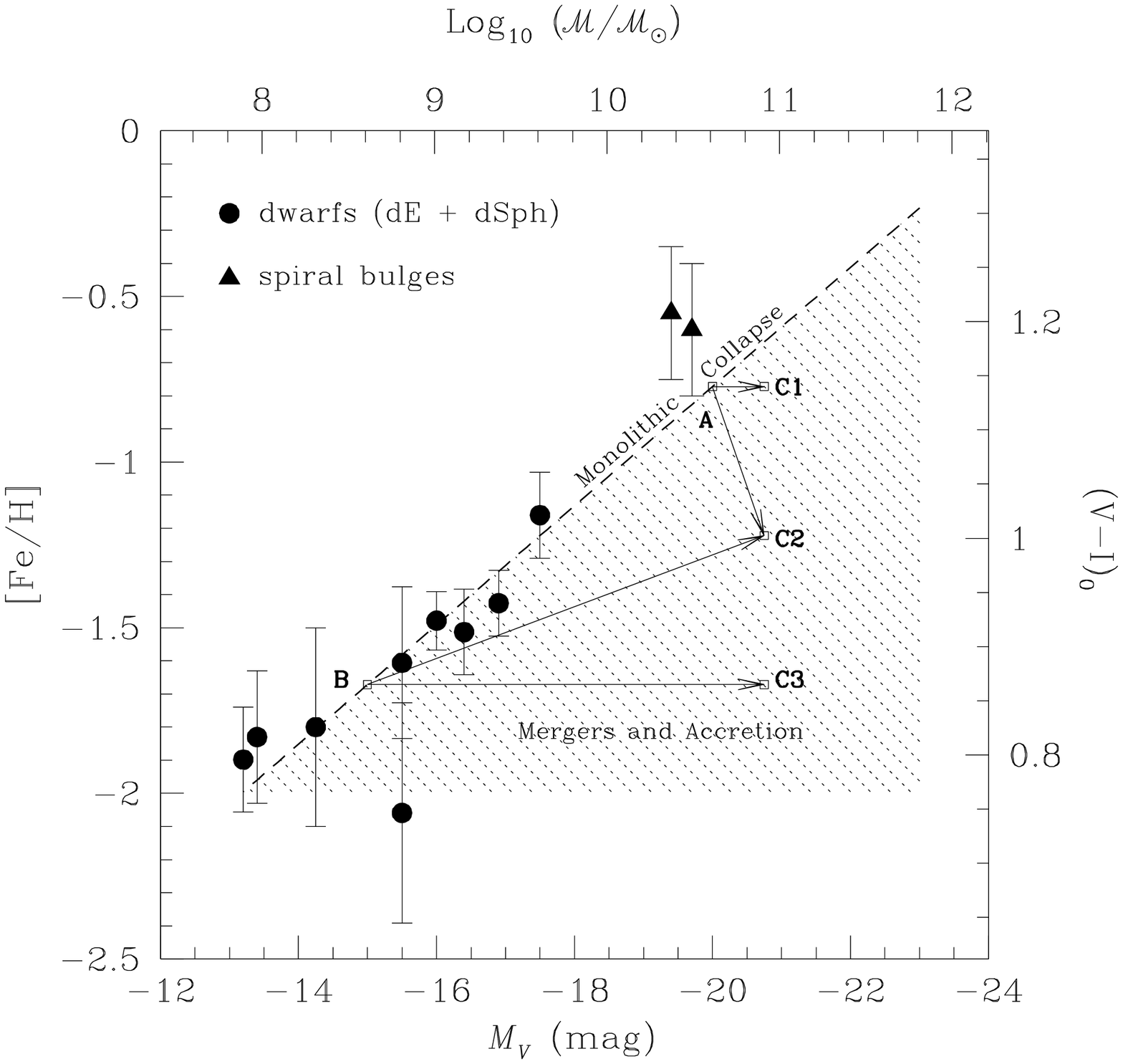}

\figcaption[ms01.eps]{
Mean color and metallicity of globular clusters belonging to low- and intermediate-luminosity 
galaxies, plotted as a function of absolute galaxy magnitude (circles). Triangles show the 
metal-rich globular cluster populations of the Milky Way and M31, plotted against their 
respective bulge luminosities. The tightness of this correlation indicates that the 
chemical enrichment of the globular clusters in these systems
was controlled primarily by the depth of the gravitational potential well 
in which they formed. The dashed line indicates our adopted color-magnitude relation 
for the GC systems of these protogalactic building blocks, each 
of which is assumed to have formed via a ``monolithic" collapse. The dotted 
region shows the area occupied by galaxies which grow hierarchically via 
dissipationless agglomeration of these protogalactic fragments.
\label{fig1}}
\clearpage

\clearpage

\plotone{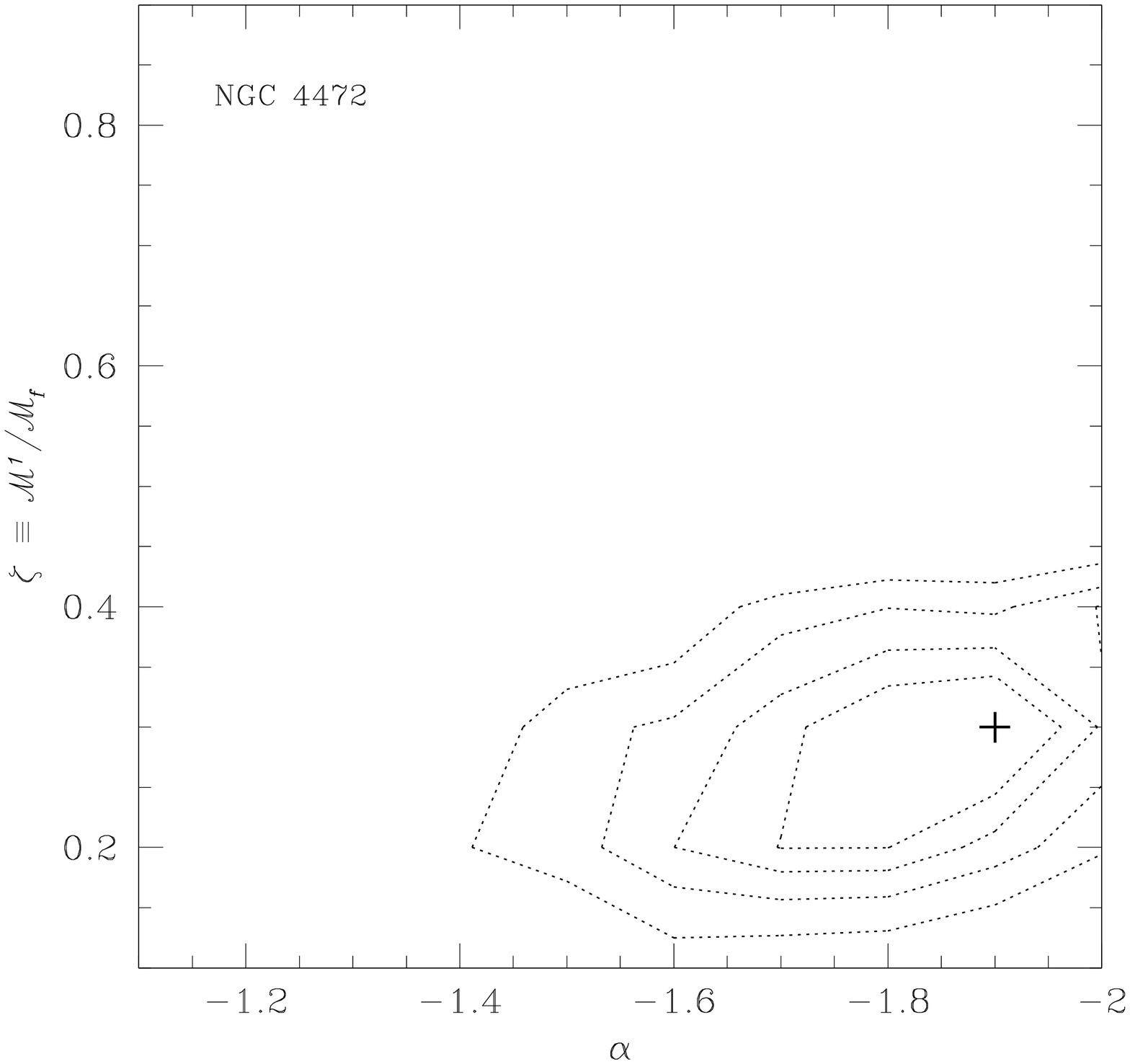}

\figcaption[ms02.eps]{Confidence intervals for $\alpha$ and $\zeta$, based
on 5,000 Monte Carlo simulations of the metallicity distribution of globular clusters 
associated with the elliptical galaxy NGC~4472. In this notation, the slope of the 
protogalactic mass/luminosity function is denoted by $\alpha$, while $\zeta$ is the ratio 
between the mass of the largest protogalactic fragment and the final mass of the galaxy. 
The contours denote 68\%, 90\%, 95\% and 99\% confidence limits.
The cross indicates the location in the $\alpha$-$\zeta$ plane of the simulation which 
produced the best match to the observed distribution. 
\label{fig2}}
\clearpage

\plotone{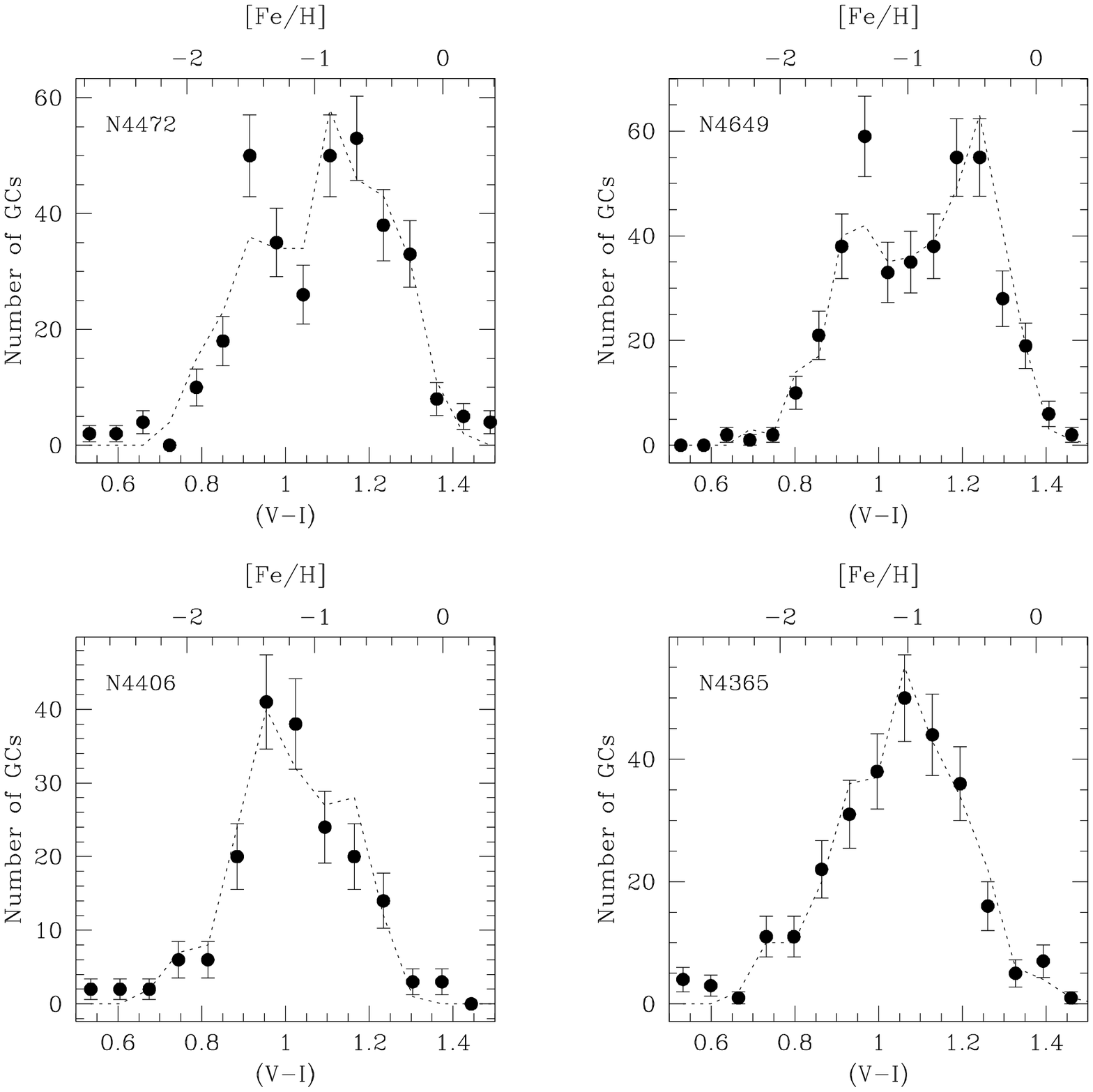}

\figcaption[ms03.eps]{
Color and metallicity distributions of globular clusters belonging to the early-type galaxies
NGC~4472, NGC~4649, NGC~4406 and NGC~4365 (filled circles). The errorbars show the Poisson 
uncertainties in each bin. The dashed curve shows the best-fit model for each galaxy, 
determined from 5,000 Monte Carlo simulations spanning a 10$\times$10 grid in ($\alpha$,~$\zeta$).
\label{fig3}}
\clearpage

\plotone{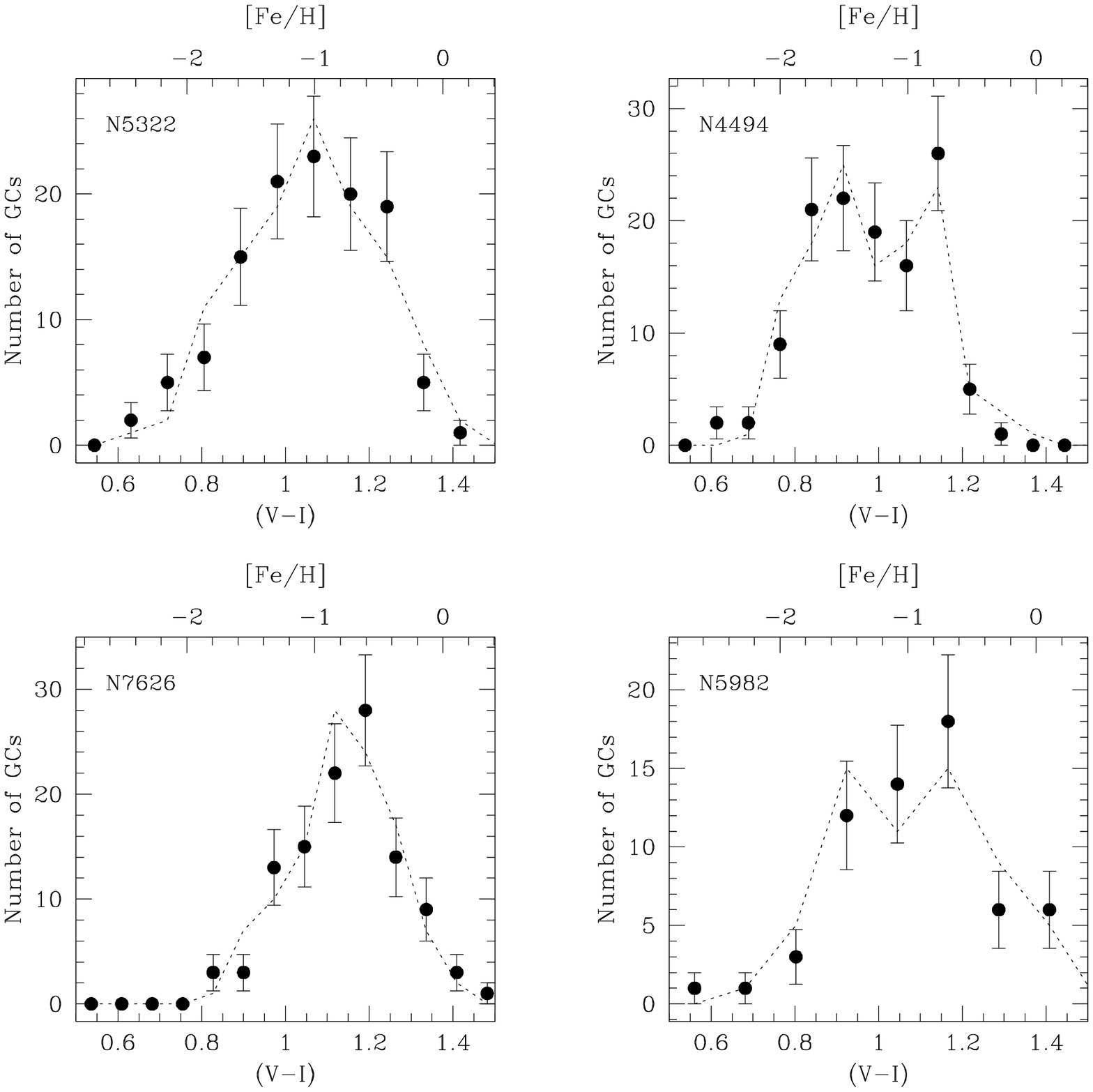}

\figcaption[ms04.eps]{
Same as in Figure~\ref{fig3}, except for NGC~5322, NGC~4494, NGC~7626 and NGC~5982.
\label{fig4}}
\clearpage

\plotone{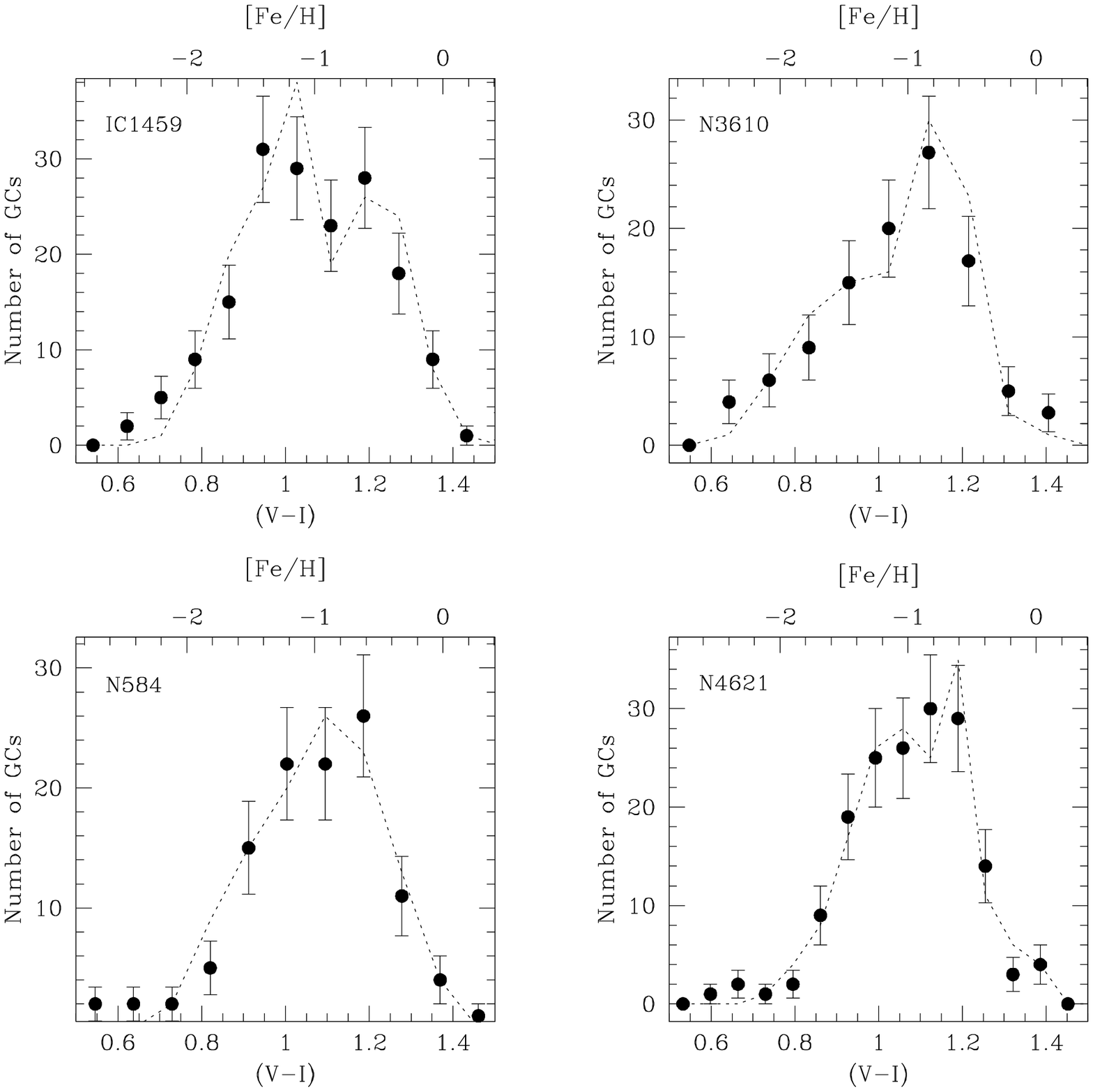}

\figcaption[ms05.eps]{
Same as in Figure~\ref{fig3}, except for IC~1459, NGC~3610, NGC~584 and NGC~4621.
\label{fig5}}
\clearpage

\plotone{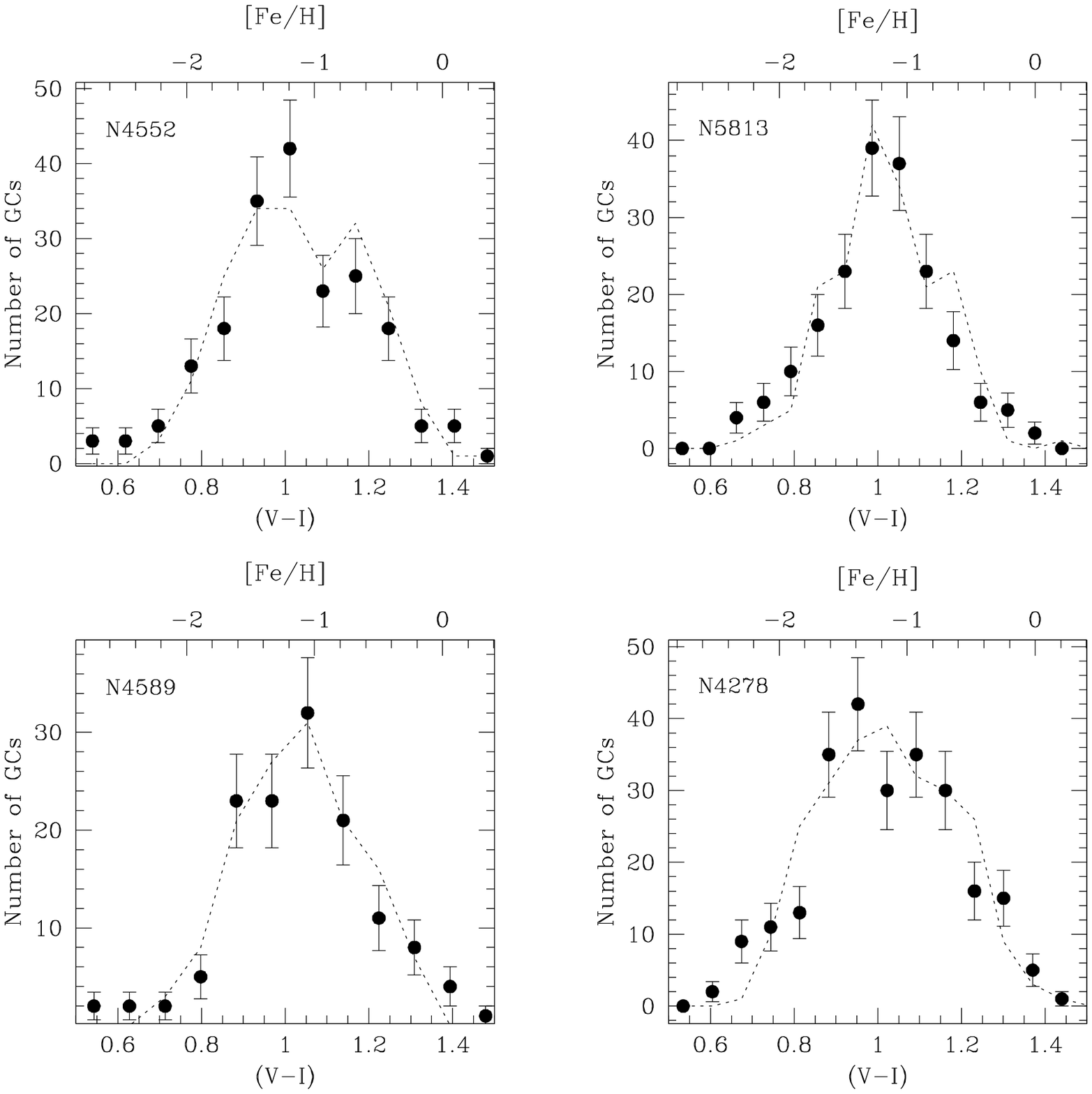}

\figcaption[ms06.eps]{
Same as in Figure~\ref{fig3}, except for NGC~4552, NGC~5813, NGC~4589 and NGC~4278.
\label{fig6}}
\clearpage

\plotone{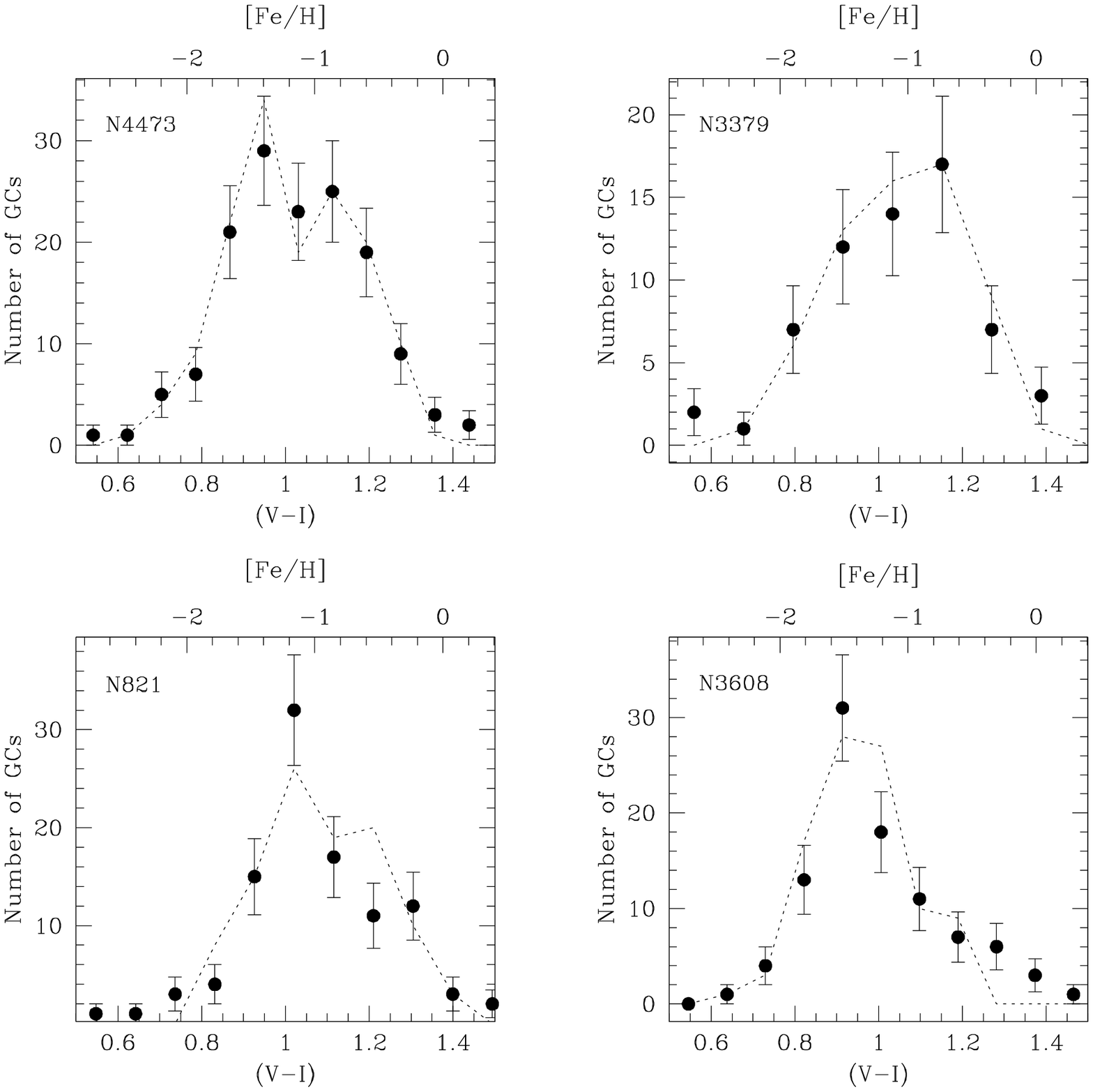}

\figcaption[ms07.eps]{
Same as in Figure~\ref{fig3}, except for NGC~4473, NGC~3379, NGC~821 and NGC~3608.
\label{fig7}}
\clearpage

\plotone{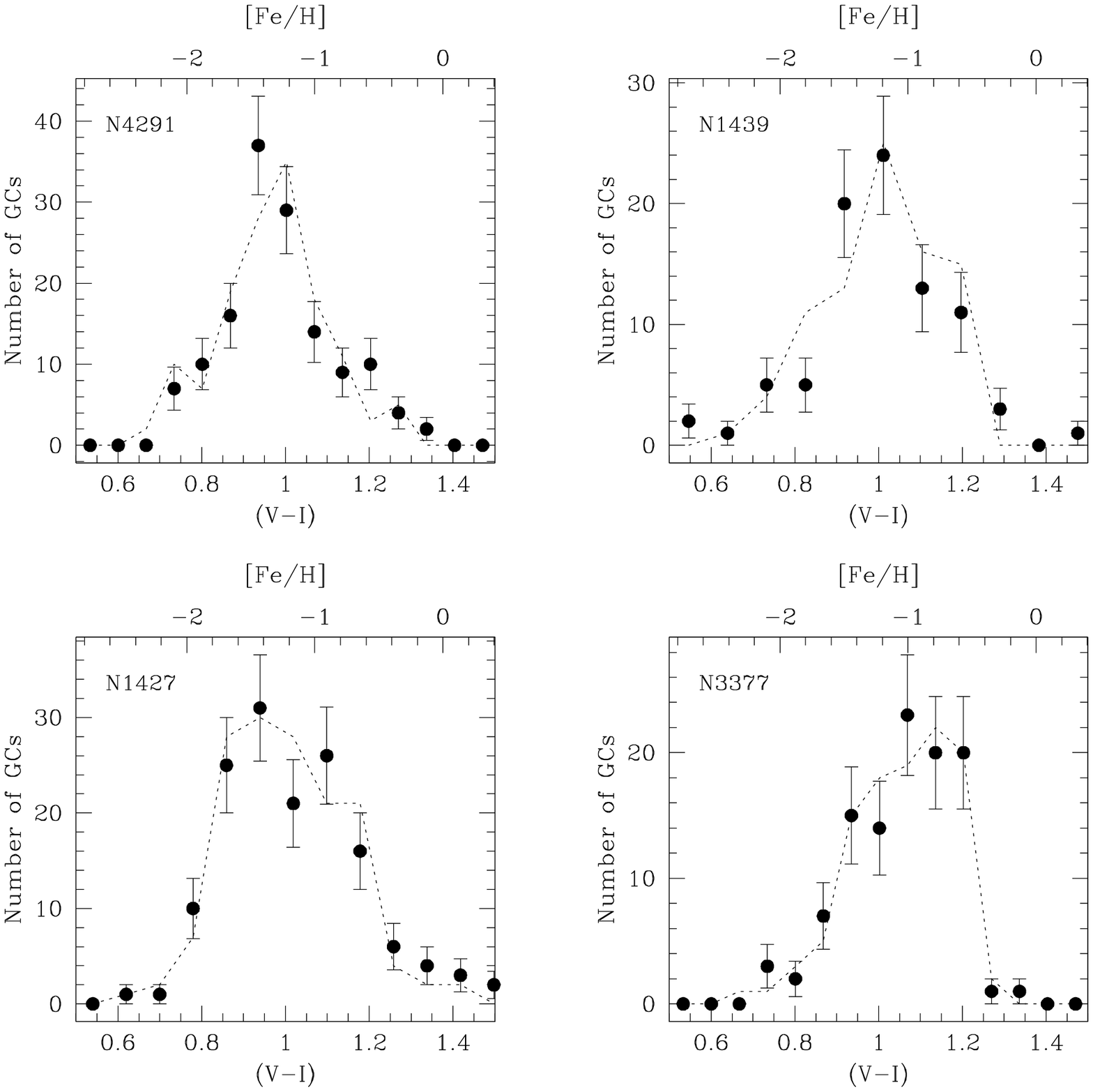}

\figcaption[ms08.eps]{
Same as in Figure~\ref{fig3}, except for NGC~4291, NGC~1439, NGC~1427 and NGC~3377.
\label{fig8}}
\clearpage

\plotone{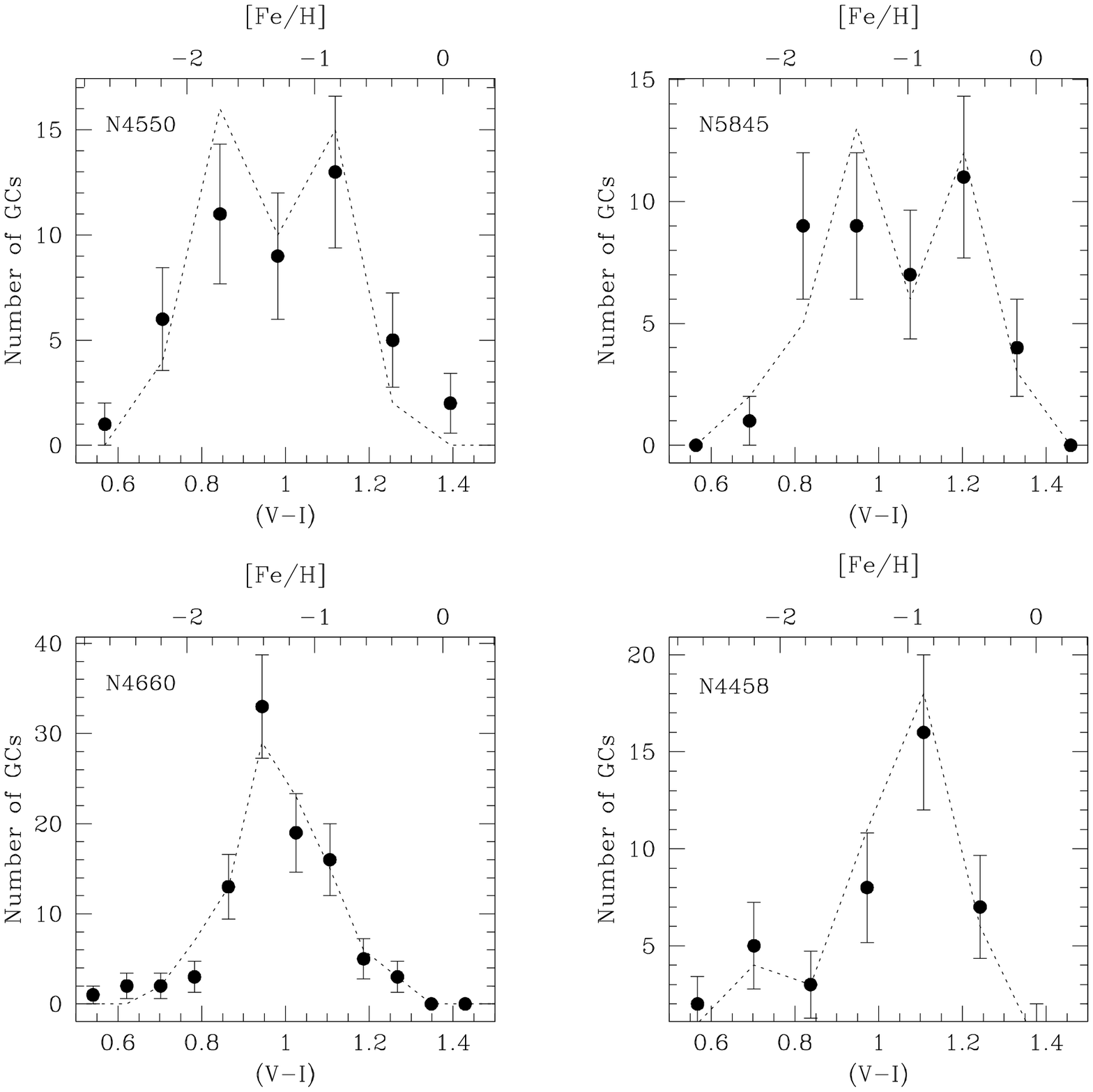}

\figcaption[ms09.eps]{
Same as in Figure~\ref{fig3}, except for NGC~4458, NGC~5845, NGC~4660 and NGC~4458.
\label{fig9}}
\clearpage

\plotone{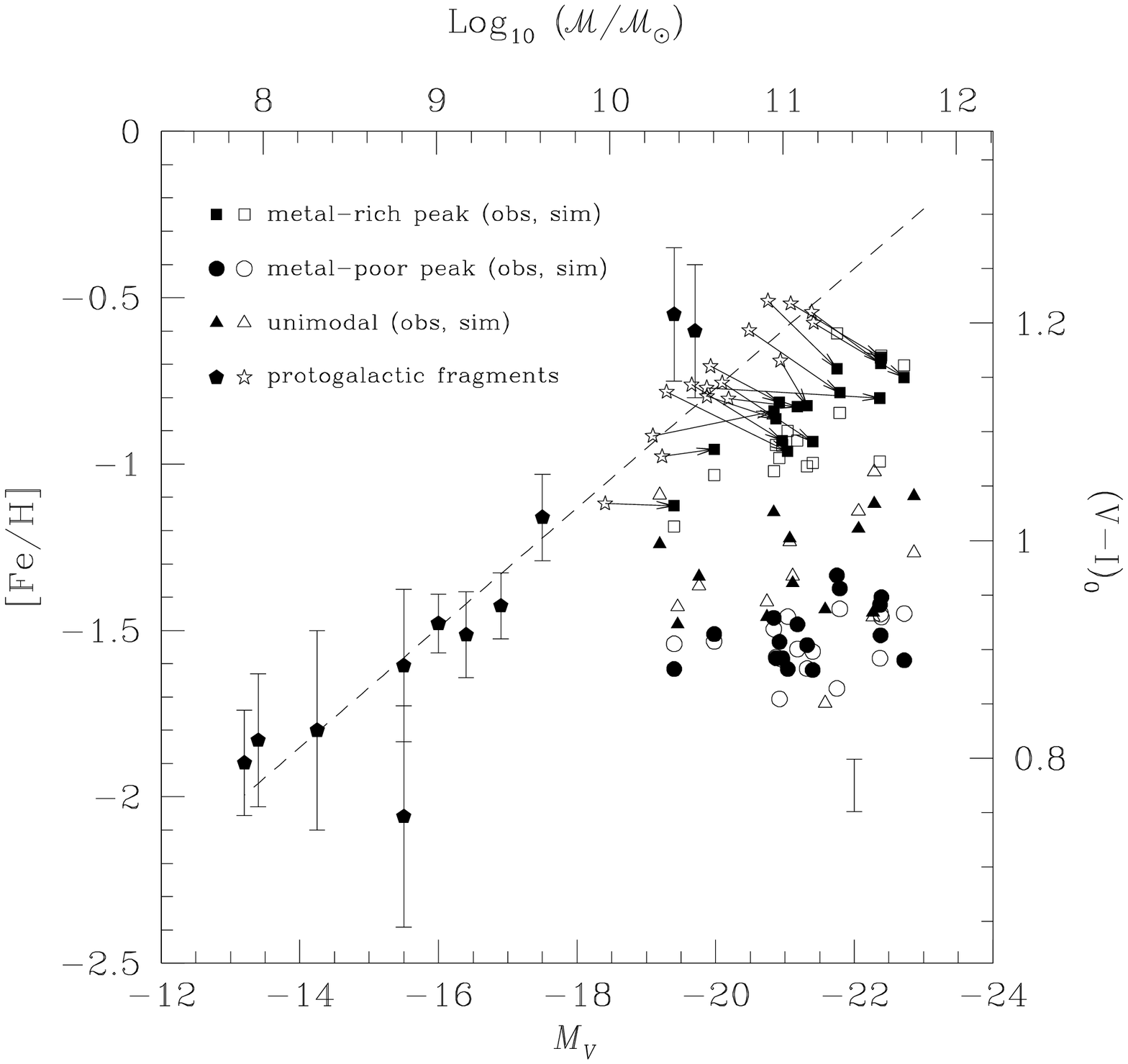}

\figcaption[ms10.eps]{
Mean color (metallicity) of globular cluster systems, plotted against galaxy magnitude 
(mass). The dashed line shows the ``zero-age" relation between galaxy mass and globular
cluster metallicity shown in Figure~\ref{fig1} for unmerged protogalactic fragments: 
$i.e.,$ isolated galaxies of low and intermediate mass, and the bulge components of 
spiral galaxies (filled pentagons). 
The filled squares and circles show the observed colors for the metal-rich and 
metal-poor globular clusters in the 16 galaxies classified as bimodal by Kundu \&
Whitmore (2001). Open squares and circles indicate the median color of these
components based on 100 simulations of the globular cluster metallicity distributions
using the ($\alpha$, $\zeta$) values reported in Table~\ref{tab1}.
The filled and open triangles show the
observed and simulated colors of globular clusters belonging to the 12 galaxies
classified by Kundu \& Whitmore (2001) as unimodal. The errorbar in the lower right corner
shows the 1$\sigma$ scatter about these median values.  For each galaxy with a
bimodal metallicity distribution, a thin arrow connects the original position of the
most massive protogalactic fragment (open stars) to the measured position
of the metal-rich peak.
\label{fig10}}
\clearpage

\plotone{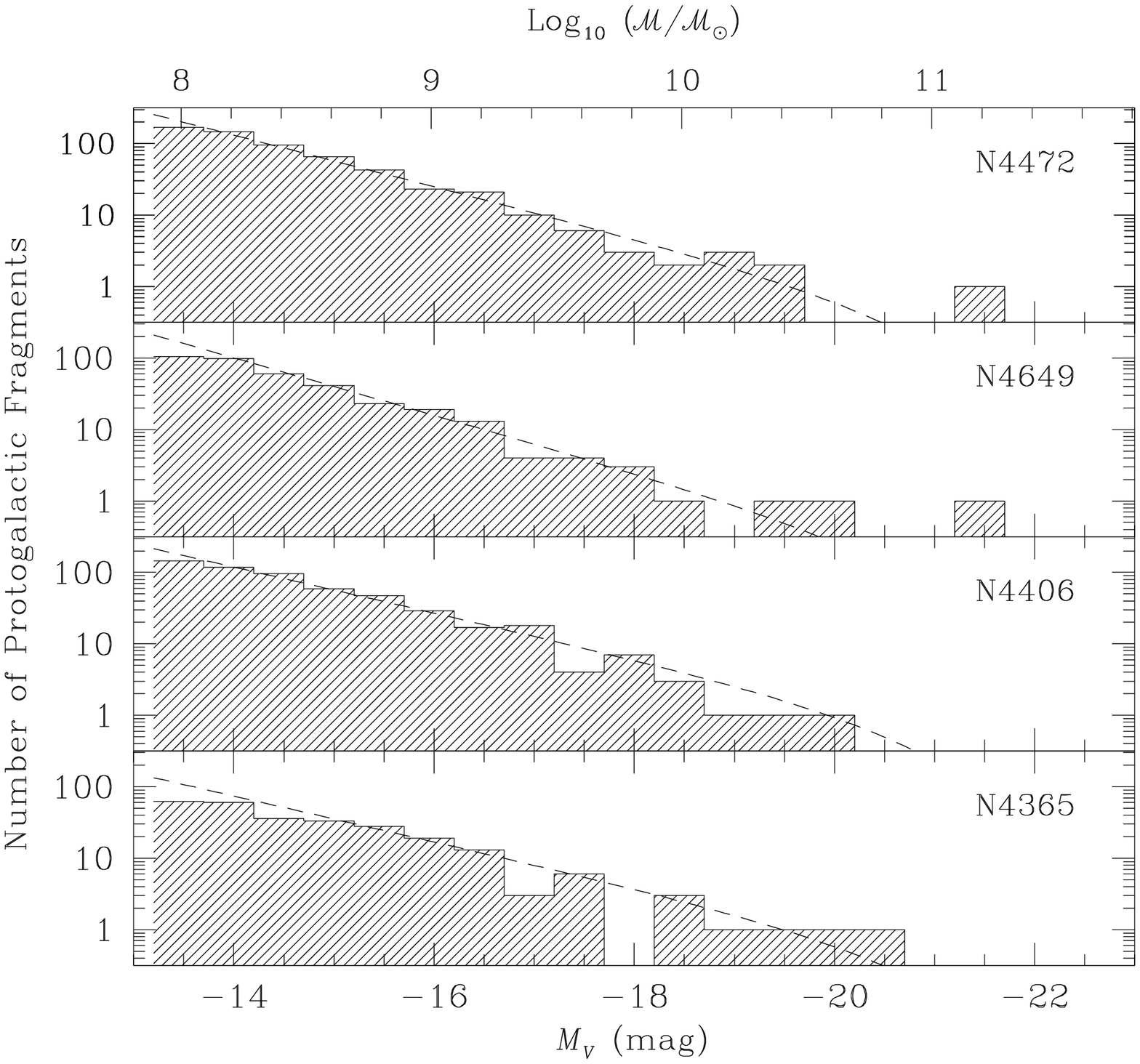}

\figcaption[ms11.eps]{
Mass and luminosity functions of protogalactic fragments for NGC~4472, NGC~4649, NGC~4406 
and NGC~4365. The histograms show the distribution of protogalactic fragments for the
best-fit models shown in Figure~\ref{fig3}. The dashed line in each panel indicates 
the parent Schechter function from which the protogalactic fragments were drawn.
\label{fig11}}
\clearpage

\plotone{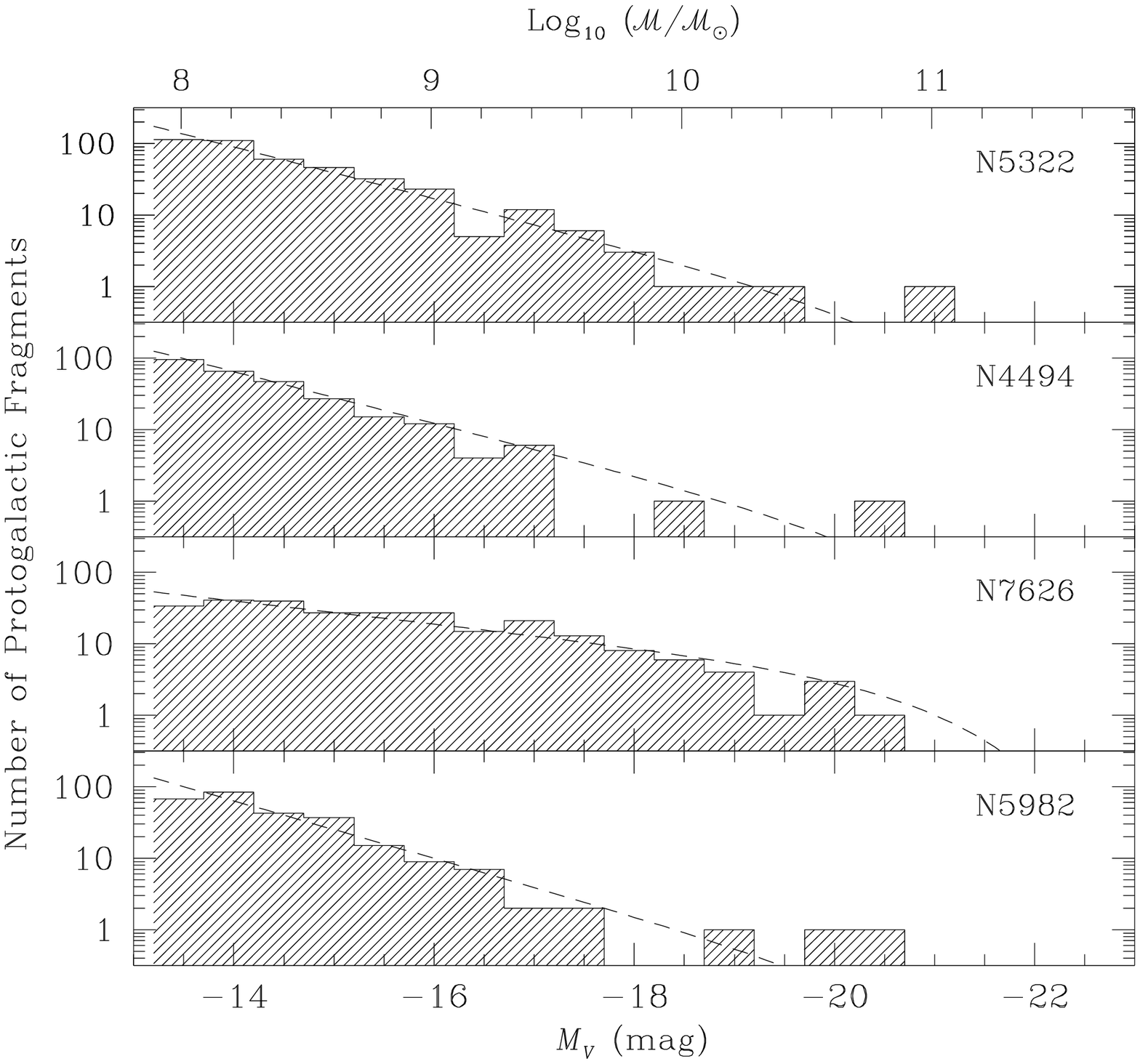}

\figcaption[ms12.eps]{
Same as in Figure~\ref{fig11}, except for NGC~5322, NGC~4494, NGC~7626 and NGC~5982.
\label{fig12}}
\clearpage

\plotone{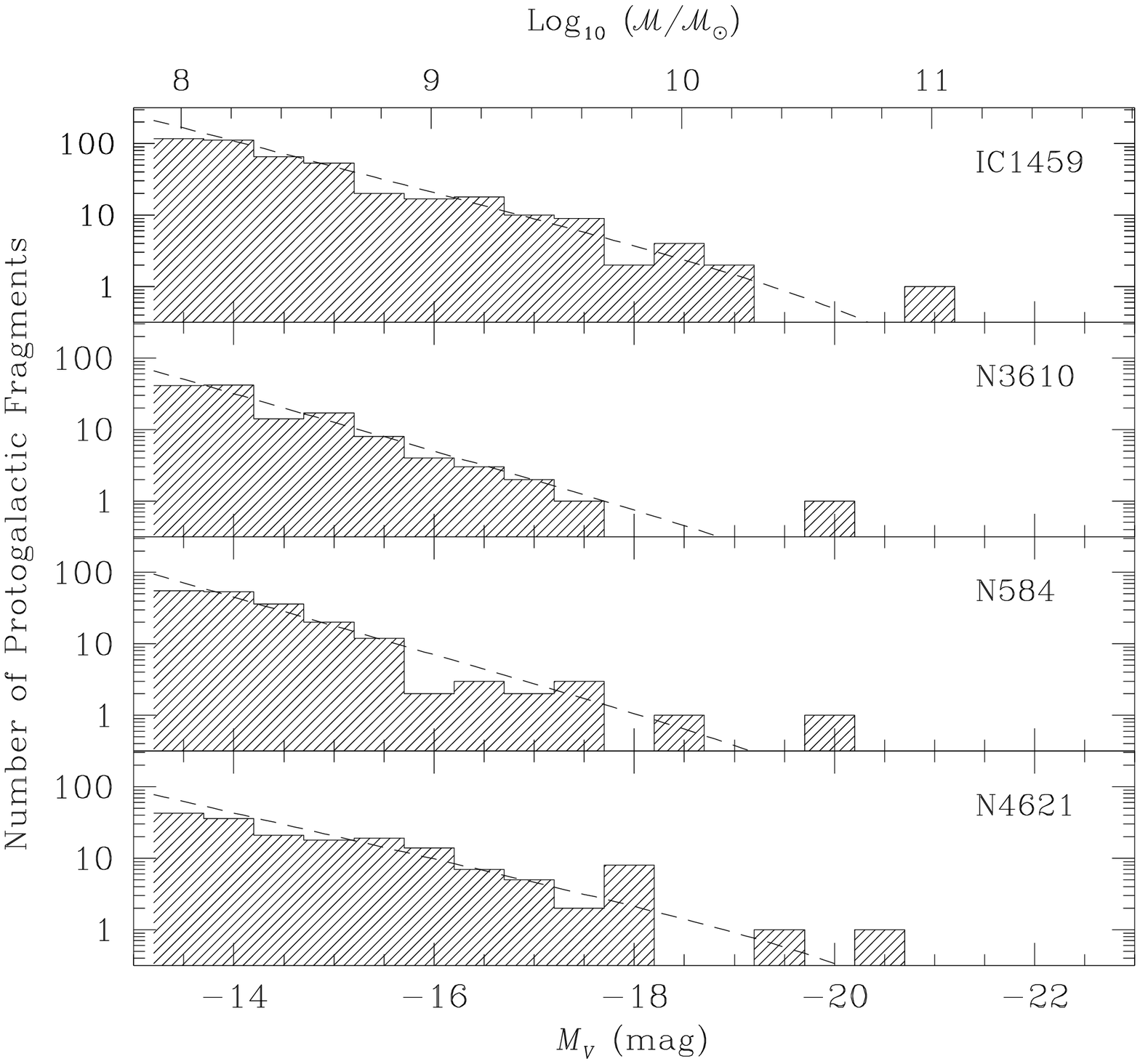}

\figcaption[ms13.eps]{
Same as in Figure~\ref{fig11}, except for IC~1459, NGC~3610, NGC~584 and NGC~4621.
\label{fig13}}
\clearpage

\plotone{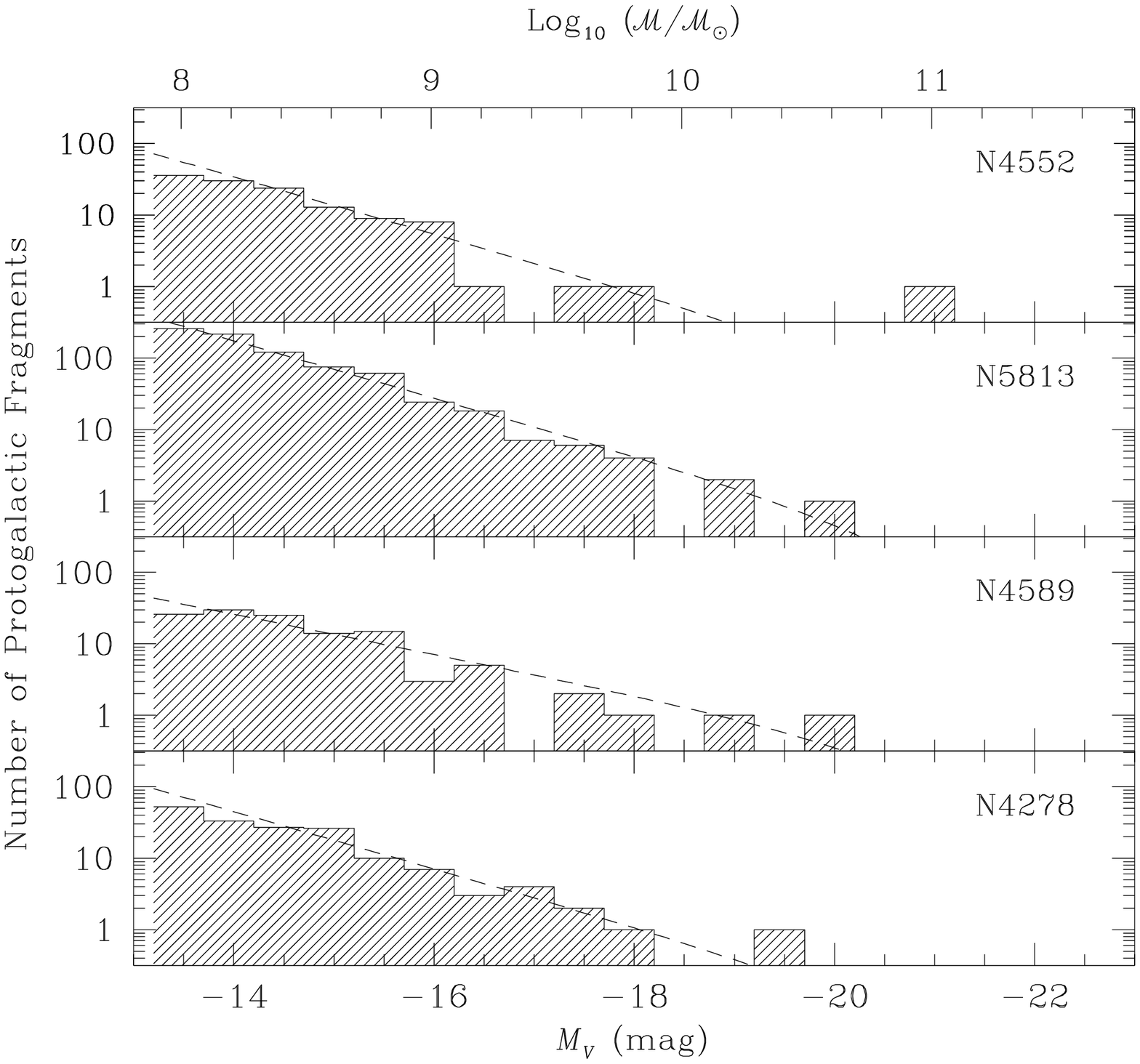}

\figcaption[ms14.eps]{
Same as in Figure~\ref{fig11}, except for NGC~4552, NGC~5813, NGC~4589 and NGC~4278.
\label{fig14}}
\clearpage

\plotone{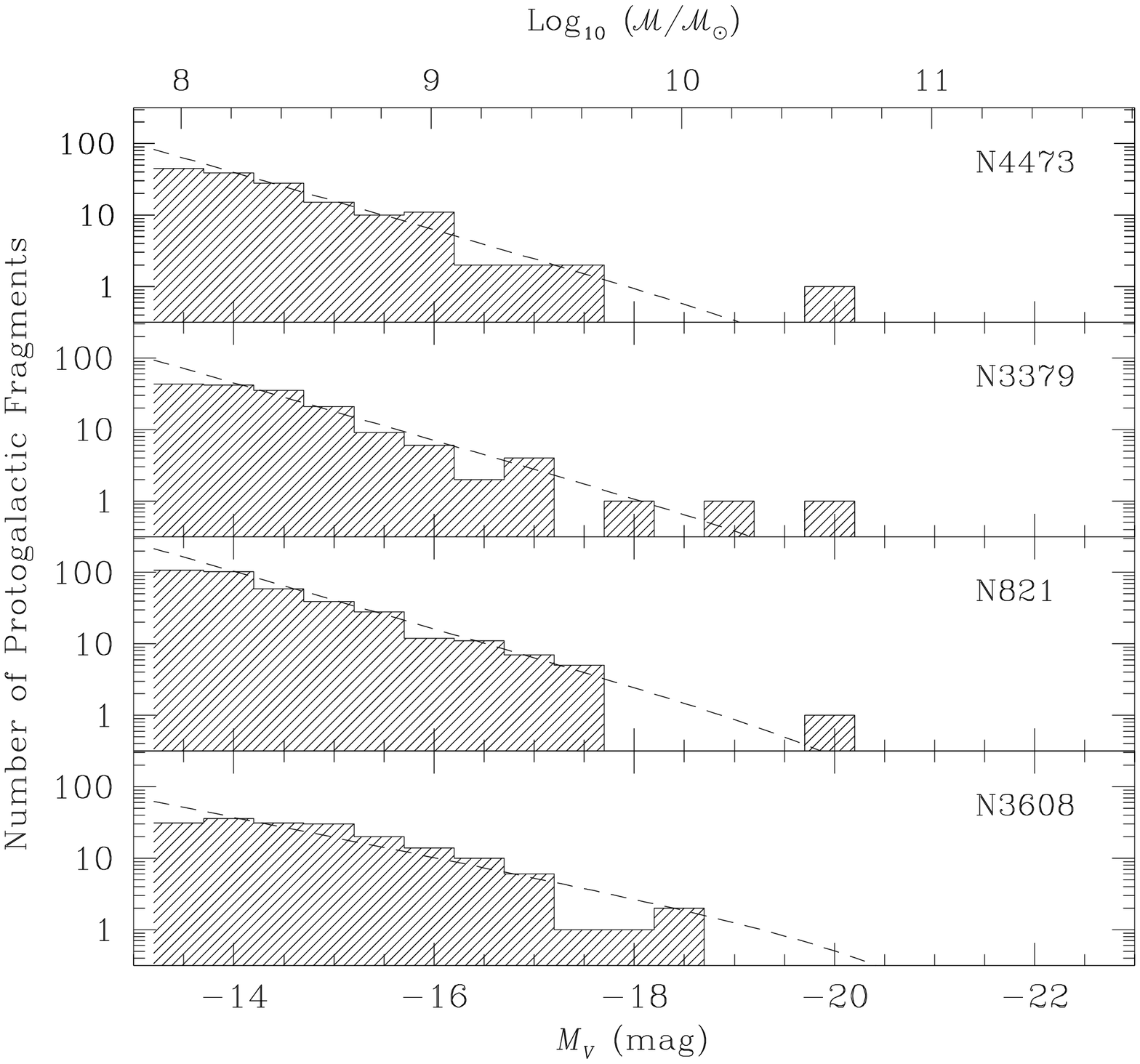}

\figcaption[ms15.eps]{
Same as in Figure~\ref{fig11}, except for NGC~4473, NGC~3379, NGC~821 and NGC~3608.
\label{fig15}}
\clearpage

\plotone{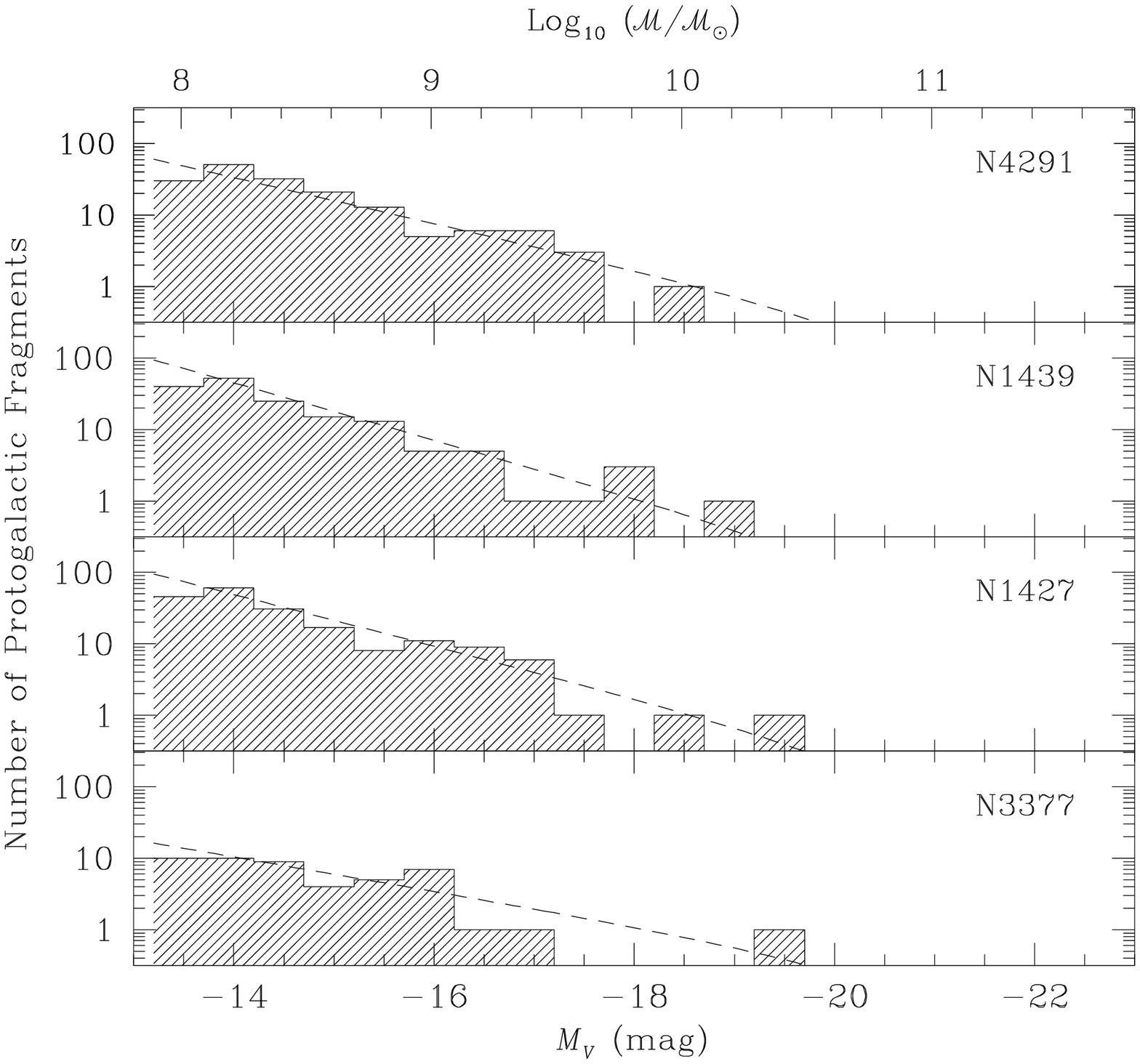}

\figcaption[ms16.eps]{
Same as in Figure~\ref{fig11}, except for NGC~4291, NGC~1439, NGC~1427 and NGC~3377.
\label{fig16}}
\clearpage

\plotone{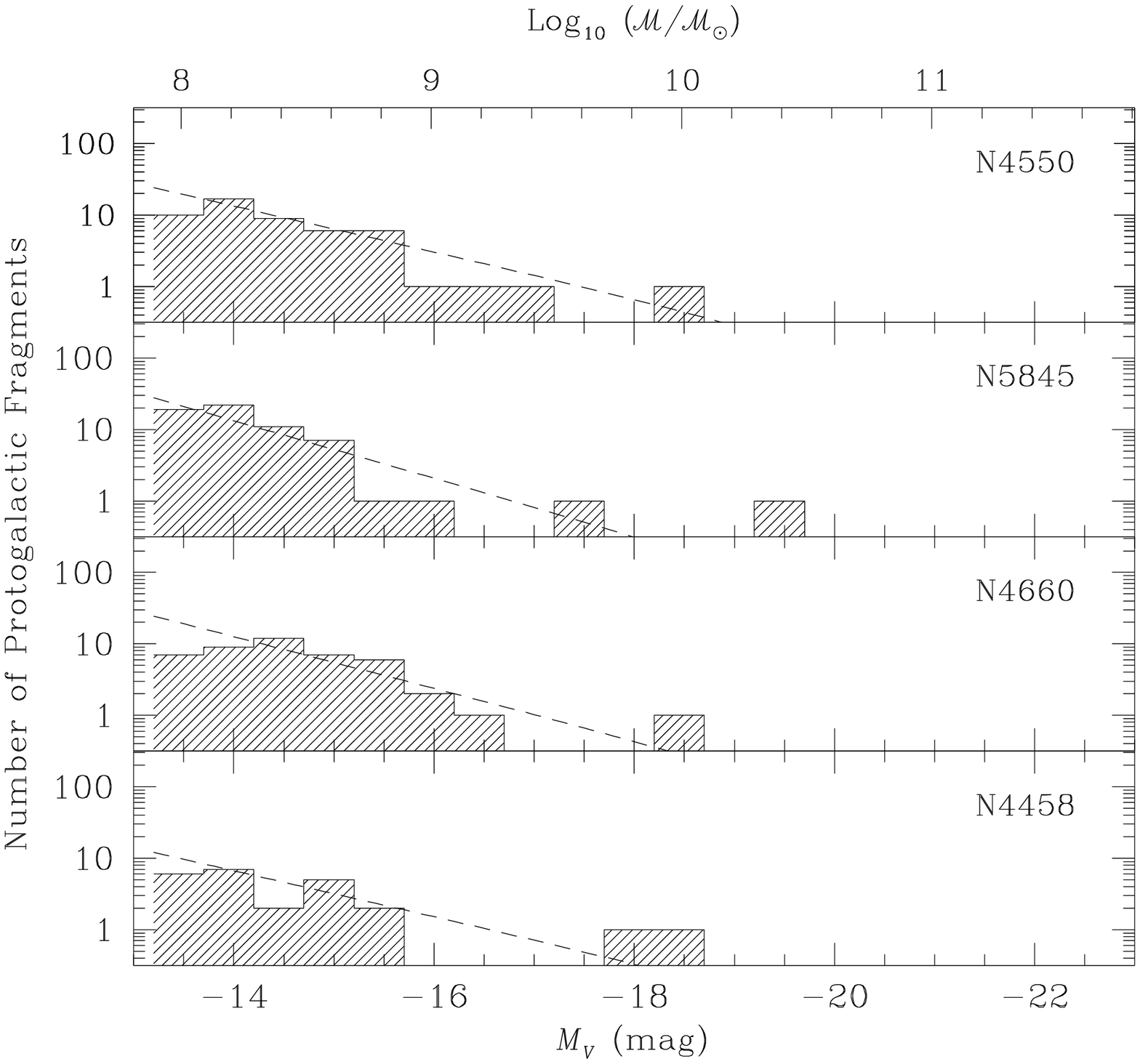}

\figcaption[ms17.eps]{
Same as in Figure~\ref{fig11}, except for NGC~4458, NGC~5845, NGC~4660 and NGC~4458.
\label{fig17}}
\clearpage

\plotone{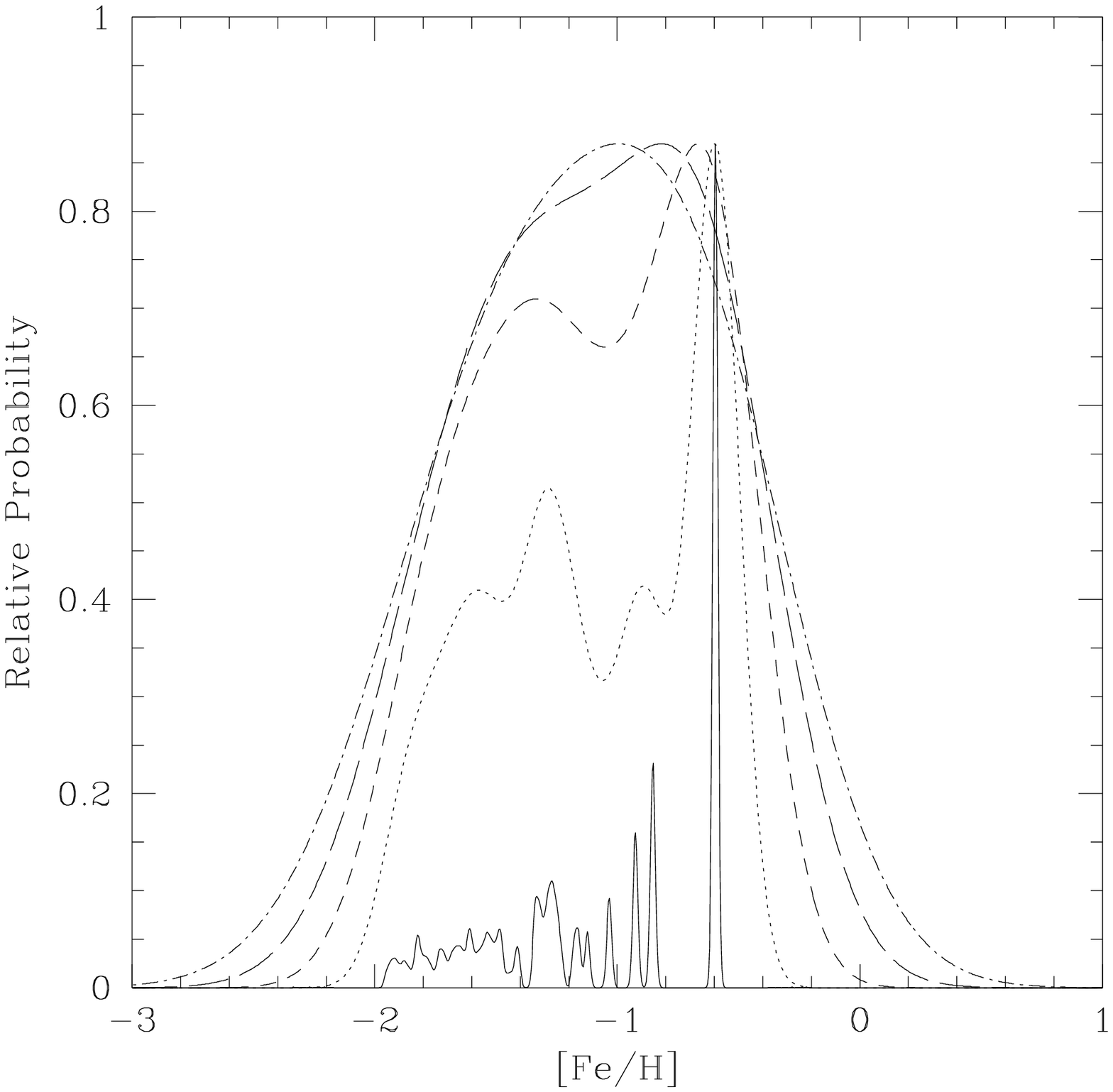}

\figcaption[ms18.eps]{
Simulated metallicity distribution for a representative galaxy, NGC 5322, assuming
the intrinsic metallicity dispersion of each protogalactic fragment is 0.01 dex (solid 
curve). The remaining curves show this same distribution after smoothing
with Gaussians having dispersions of 0.1, 0.2, 0.3 and 0.4 dex. 
\label{fig18}}

\plotone{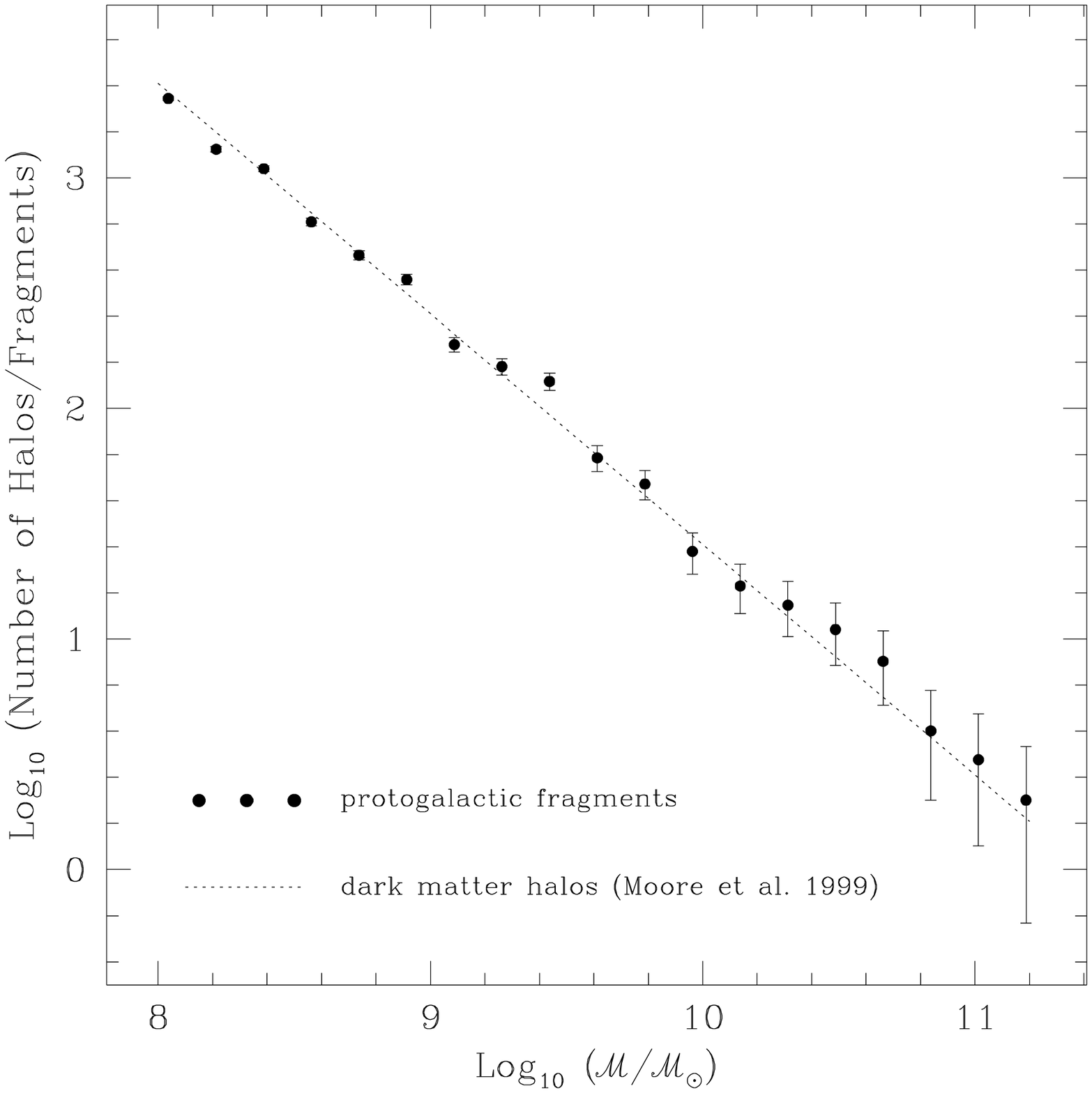}

\figcaption[ms19.eps]{
Comparison of the halo mass function measured from cosmological N-body simulations 
(dotted line) with the mass spectrum of protogalactic fragments inferred from the 28 
globular cluster systems studied here (filled circles). Following Moore et al. (1999), 
we express the mass function of dark matter halos as $n({\cal M}) \propto {\cal M}^{-2}$.
\label{fig19}}

\end{document}